\documentstyle[preprint,eqsecnum,aps,amsfonts]{revtex}
\tightenlines

\newcommand{\be}{\begin{equation}}
\newcommand{\ee}{\end{equation}}
\newcommand{\ba}{\begin{array}}
\newcommand{\ea}{\end{array}}
\newcommand{\bea}{\begin{eqnarray}}
\newcommand{\eea}{\end{eqnarray}}

\newcommand{\lra}{\longrightarrow}
\newcommand{\mb}{\mbox}

\newcommand{\ve}{\varepsilon}
\newcommand{\T}{\textstyle}
\newcommand{\Sc}{\scriptscriptstyle}
\newcommand{\D}{\displaystyle}
\newcommand{\tr}{\mbox{tr}}
\newcommand{\diag}{\mbox{diag}}
\newtheorem{claim}{Solution}[section]
\newtheorem{proposition}{Proposition}

\begin{document}
\title{SU(3) density matrix theory}
\author{Ts. Dankova and G. Rosensteel}
\address{
Department of Physics, Tulane University\\
New Orleans, LA 70118 USA}
\date{\today}
\maketitle
\begin{abstract}
The irreducible representations of the Lie algebra ${\frak su}$(3) describe rotational bands in the context of the nuclear shell and interacting boson models. The density matrices associated with ${\frak su}$(3) provide an alternative theoretical framework for obtaining these bands. The ${\frak su}$(3) density matrix formulation is mathematically simpler than representation theory, yet it yields similar results. Bands are solutions to a system of polynomial equations defined by the quadratic and cubic ${\frak su}$(3) Casimirs. Analytic solutions are found in many physically important cases including rotation about principal axes and spheroids. Numerical solutions are reported in other cases including tilted rotors. The physics of ${\frak su}$(3) rotational bands is more transparent in the density formalism than in representation theory.
\end{abstract}

\narrowtext

\section{Introduction}
\label{sec:level1}

The physics of many-particle systems is dominated in many cases by a relatively small number of degrees of freedom.  The quintessence of the physics is clarified by models which ignore minor effects and focus on the principal dynamical factors. When the set of relevant observables closes under commutation to form a Lie algebra, it is called a spectrum generating or dynamical symmetry algebra \cite{dothan65,dothan70,joseph74}, and a group theoretical model is suggested as the natural explanatory framework.

If ${\frak g}$ is a spectrum generating Lie algebra of hermitian operators, then its irreducible unitary representations (irreps) define group theoretical models. The decomposition of the reducible representation of ${\frak g}$ on Fock space into its irreducible subspaces provides the microscopic interpretation of the models. Although an irrep of ${\frak g}$ simplifies the original Fock space problem, its dimension may be still too large to allow an easy analysis of the physics, e.g., noncompact algebras like ${\frak sp}(3,{\Bbb R})$ of the symplectic collective model \cite{ROS77a,ros80,rowe96} or ${\frak u}(6,6)$ of the extended interacting boson model \cite{decoster96} have infinite-dimensional irreducible representations. Intractable models also may arise when the rotation group algebra ${\frak so}(3)$ is not canonically embedded in ${\frak g}$ and the angular momentum decomposition is not multiplicity free, e.g., the compact symplectic group ${\frak sp}(2j+1)$ which describes seniority in a single $j$-shell \cite{racah42,talmi91}, or the $sdg$-boson model based on ${\frak u}(15)$ \cite{wu84,isacker81}.

The relationships among the shell model, Hartree-Fock, and the group $U(n)$ of unitary transformations in the $n$-dimensional single-particle space suggest a solution \cite{RR81,R81,Kramer}.

The shell model may be regarded as a group theoretical model in which the spectrum generating algebra is the set of all one-body hermitian operators ${\frak u}(n)$. Throughout this article, Lie algebras are denoted by ${\frak g}$, e.g., ${\frak u}(n)$, ${\frak su}(3)$, ${\frak so}(3)$, while the corresponding Lie group  is written as $G$, e.g., $U(n)$, $SU(3)$, $SO(3)$. Each Lie group is the exponentiation of its Lie algebra, e.g., $U(n) = \exp({\frak u}(n))$. For $k$ identical valence fermions, the shell model space is the totally antisymmetric irreducible representation of $U(n)$ with dimension $n!/[k!(n-k)!]$. Already for medium mass nuclei with active valence neutrons and protons, the factorially-growing dimension of the antisymmetric irrep is astronomical. Moreover the rotation group is not canonically embedded in $U(n)$.

A way around this difficulty is self-consistent mean field theory. In Hartree-Fock, the mean field Hamiltonian must be diagonalized in the single-particle space of dimension $n$ -- no matter how many particles are in the valence space.  The set of admissible states in the Hartree-Fock approximation is the set of Slater determinants, or, equivalently, the set of idempotent hermitian density matrices, $\rho^2 = \rho$, with trace equal to the number $k$ of valence nucleons, $\mbox{tr}\, \rho = k$. The density matrices are defined by the expectation of the one-body hermitian operators in ${\frak u}(n)$. Mathematically, the set of density matrices spans the dual space of the algebra ${\frak u}(n)$. The set of admissible idempotent densities is not a vector space, but a surface of dimension $2k(n-k)$ contained within the vector space of all hermitian density matrices \cite{RR81}. 

The surface of idempotent densities is a level surface of the Casimirs. There are $n$ Casimirs for the unitary algebra ${\frak u}(n)$ given by the trace of powers of the density, ${\cal C}_r = \mbox{tr}\, \rho^r$ for $r=1,\ldots, n$. On the idempotent density surface, the Casimirs are evidently constant ${\cal C}_r =  k$ for all $r$. Conversely, the level surface in the dual space of the Casimirs ${\cal C}_r =  k$ for integral $k\in[0,n]$ consists of the Hartree-Fock densities.

The idempotent densities transform among themselves by the elements of the unitary group: a unitary matrix $g$  transforms a density $\rho$ into $g \rho g^{-1}$. $U(n)$ is a transformation group on each level surface of the Casimirs.

Thus there is a common algebraic structure for both the shell model and Hartree-Fock. The distinctions arise from the way the algebra determines the model states. For the shell model, the quantum states span an irreducible unitary representation of the group $U(n)$. For Hartree-Fock, the role of the unitary group and its Lie algebra has three essential aspects: (1) the densities in mean field theory are elements of the dual space of the Lie algebra, (2) the allowed Hartree-Fock densities are a level surface for the Casimirs, and (3) the unitary group transforms the allowed densities among themselves. These are the three ingredients for a mean field theory that may be implemented for any Lie algebra.

The aim of this article is to construct a density matrix or mean field theory for the test case of the $SU(3)$ model that underlies the algebraic theory of nuclear rotational motion. $SU(3)$ is a paradigm algebraic theory since it is the simplest nontrivial dynamical symmetry in nuclear structure physics. The relationship between $SU(3)$ irreducible representations and their mean field approximations is investigated. It is shown that there is a close correspondence between the results from these two models. But the $SU(3)$ density matrix theory is mathematically simpler and provides a clear physical interpretation of the rotational bands found in the irreps.

\section{${\frak su}$(3) density matrices}
\label{sec:level2}
\subsection{Algebra definition}
Let $(x_{\alpha j}, p_{\alpha j})$ denote the Cartesian components of the dimensionless position and momentum hermitian operators of particle $\alpha$ in a system of $k$ particles. They obey the canonical commutation relation $[x_{\alpha j}, p_{\beta k}]=i\delta_{\alpha \beta} \delta_{jk}$. The traceless Elliott quadrupole operator \cite{Elliott} 
\be \label{quaddef}
\hat{\cal Q}^{(2)}_{jk} = \T{\frac{1}{2}}\left(x_{\alpha j} x_{\alpha k} + p_{\alpha j} p_{\alpha k}- {\T{\frac{1}{3}}}\delta_{jk}(x_{\alpha\mu} x_{\alpha\mu} +p_{\alpha\mu} p_{\alpha\mu})\right)
\ee
and the vector angular momentum operator
\bea
\hat{L}_{jk} & = & x_{\alpha j} p_{\alpha k} - x_{\alpha k} p_{\alpha j} \nonumber \\
\hat{L}_i & = & \T{\frac{1}{2}}\, \ve_{ijk} \hat{L}_{jk}
\eea
(summation over repeated indices) generate an eight-dimensional real Lie algebra of one-body hermitian operators,
\be
\ba{rcl}
\big[\,\hat{L}_j\,,\hat{L}_k\big] &=& i\,\ve_{jkm}\hat{L}_m \\ 
\big[\hat{\cal Q}^{\Sc (2)}_{jk}\,,\hat{L}_r\big] &=& i\,(\ve_{rsj}\hat{\cal Q}^{\Sc (2)}_{sk}  + \ve_{rsk}\hat{\cal Q}^{\Sc (2)}_{sj})\\
\big[\hat{\cal Q}^{\Sc (2)}_{jk}\,,\hat{\cal Q}^{\Sc (2)}_{rs}\big] &=&i\,\T{\frac{1}{4}}\left( \delta_{kr}\ve_{js\alpha}\hat{L}_\alpha + \delta_{ks}\ve_{jr\alpha}\hat{L}_\alpha \right. \nonumber \\
& & \left. + \delta_{jr}\ve_{ks\alpha}\hat{L}_\alpha + \delta_{js}\ve_{kr\alpha}\hat{L}_\alpha \right)\,, 
\ea
\ee
that is isomorphic to the algebra of hermitian traceless matrices
\be
{\frak su}(3)= \left \{ Z \in M_3({\Bbb C}) \, \left| \, Z^{\dagger}=Z \,,\, \tr Z=0 \right. \right \}.
\ee
If $X$ and $Y$ are real $3\times 3$ matrices, $Z=Y+iX \in {\frak su}(3)$
if and only if $X^{\Sc T}=-X$, $Y^{\Sc T}=Y$, and $\tr\,Y=0$. The isomorphism $\sigma$ between the algebra of matrices and the algebra of hermitian operators is given explicitly by:
\be
\sigma(Z)=Y_{jk}\hat{\cal Q}^{\Sc (2)}_{jk}-{\T \frac{1}{2}}\, X_{jk}\hat{L}_{jk} 
\ee
for $Z=Y+iX\in{\frak su}(3)$. Note that $[\sigma(Z), \sigma(W)] = i\,\sigma([Z,W])$ for $Z,W\in {\frak su}(3)$.

Alternatively, the algebra of hermitian operators may be defined as the ${\frak su}$(3) dynamical symmetry algebra of the interacting boson model \cite{Arima78,iachello,ros90}. If $d^\dagger$ and $s^\dagger$ denote the boson creation operators, then the generators are the angular momentum and quadrupole operator,
\bea
\hat{L}_\mu & = & \sqrt{10} [ d^\dagger \times \tilde{d} ]^{(1)}_\mu \nonumber \\
\hat{\cal Q}^{(2)}_\mu & = & d^\dagger_\mu s + s^\dagger \tilde{d}_\mu \mp \frac{\sqrt{7}}{2} [ d^\dagger \times \tilde{d} ]^{(2)}_\mu ,
\eea
where $\mp$ corresponds to particle and hole bosons, respectively. This IBM algebra of hermitian operators is also isomorphic to the algebra of matrices ${\frak su}$(3).

Mathematically, all isomorphic copies of an algebra are indistinguishable. The ${\frak su}$(3) algebra of matrices is more convenient to use for calculations than either the Elliott or IBM operator algebras.

\subsection{Dual space}
Consider an  algebra of hermitian operators acting on the Hilbert space ${\cal H}$ which is isomorphic to ${\frak su}(3)$, e.g., the Elliott or IBM representations. For each normalized state vector $\Psi \in {\cal H}$ the expectations of the operators
\bea
q_{jk} & = & \langle \Psi \mid \hat{Q}^{(2)}_{jk} \mid \Psi \rangle \nonumber \\
l_{jk} & = & \langle \Psi \mid \hat{L}_{jk} \mid \Psi \rangle .
\eea
define a real symmetric traceless matrix $q$ and a real antisymmetric matrix $l$. The ``density" matrix corresponding to $\Psi$ is defined as the hermitian traceless matrix $\rho=q-\frac{1}{2}\,i\,l$. In terms of it, the expectation of a general element of the operator algebra is
\be
\langle \rho\,, Z \rangle = \mbox{tr}(\rho Z) = \langle \Psi \mid \sigma(Z) \mid \Psi \rangle,
\ee
for $Z=Y+iX \in {\frak su}(3)$.

In fact each traceless hermitian density matrix $\rho$ defines a real-valued linear functional on the matrix Lie algebra ${\frak su}(3)$, viz., $\langle \rho\,, Z \rangle = \mbox{tr}(\rho Z)$ for all $Z\in{\frak su}(3)$. The set of all such linear functionals is called the dual space of  ${\frak su}(3)$ and is denoted by ${\frak su}(3)^\ast$. In Dirac quantum mechanics the dual space is the space of ``bras." The mapping from the Hilbert space to the dual space is called the moment map $M : {\cal H} \rightarrow {\frak su}(3)^\ast$ where the density corresponding to the  vector $\Psi$ is $\rho = M(\Psi)$ \cite{Souriaufr,Kostant,Marsden}.

The density retains only part of the entire information about the system that the wave function carries, but a very important part -- the expectations of the ${\frak su}(3)$ observables. The dimension of the dual space is the same as the dimension of the ${\frak su}(3)$ algebra and is a significant simplification of the quantum problem in the Hilbert space ${\cal H}$. It reduces all the degrees of freedom incorporated in the wave function to just those most relevant to the physics of ${\frak su}(3)$ rotational states.

When $\Psi_{\Sc{HW}}$ is a highest weight vector for an irreducible representation of ${\frak su}(3)$, the corresponding density is a diagonal matrix. To see this, express the Elliott generators in terms of  the one-body operators
\be
\hat{C}_{jk}= \frac{1}{2} \left( a^{\dagger}_{\alpha j}a^{}_{\alpha k} + a^{}_{\alpha k}a^{\dagger}_{\alpha j} \right) ,
\ee
where $a^{}_{\alpha k}$, $a^{\dagger}_{\alpha k}$ are the harmonic oscillator bosons, $a_{\alpha k} = (x_{\alpha k} + i\,p_{\alpha k})/\sqrt{2}$, $a^{\dagger}_{\alpha k} = (x_{\alpha k} - i\,p_{\alpha k})/\sqrt{2}$.
The Elliott ${\frak su}(3)$ generators are given by
\bea
\hat{\cal Q}^{(2)}_{jk} & = &  \frac{1}{2}\left(\hat{C}_{jk}+\hat{C}_{kj}-\frac{2}{3}\delta_{jk}H_0\right) \nonumber \\
\hat{L}_{jk} & = & -i\, \left(\hat{C}_{jk} - \hat{C}_{kj}\right) ,
\eea
where $H_0$ is the harmonic oscillator Hamiltonian, $H_0 = \hat{C}_{jj}$.
By definition, the highest weight state is annihilated by the raising operators and is an eigenvector of the ${\frak su}(3)$ Cartan subalgebra,
\bea
\hat{C}_{jk}\Psi_{\Sc{HW}}  & = & 0, \mbox{\ when\ }j<k \nonumber \\
(\hat{C}_{33}-\hat{C}_{11})\Psi_{\Sc{HW}}  & = & \lambda\, \Psi_{\Sc{HW}}  \\
(\hat{C}_{11}-\hat{C}_{22})\Psi_{\Sc{HW}}  & = & \mu\, \Psi_{\Sc{HW}} \nonumber,
\eea
when the weights $\lambda, \mu$ are nonnegative integers. Therefore the density of the highest weight state is
\be 
\rho_{(\lambda \mu)} = \frac{1}{3} \left ( \ba{ccc} -\lambda + \mu  & 0 & 0 \\ 0 &  - \lambda - 2\mu  &  0 \\ 0	& 0 &  2\lambda + \mu \\
\ea \right )\,. \label{matrixrho}
\ee
A similar argument for the IBM ${\frak su}$(3) algebra yields the same diagonal density matrix for the highest weight vector. Indeed the derivation is independent of the specific realization of the ${\frak su}$(3) operator algebra.

\subsection{SU(3) group transformation}
The group $SU(3)$ consists of the complex $3\times 3$ unitary matrices with unit determinant.  By exponentiation, a representation $\sigma$ of the Lie algebra ${\frak su}(3)$ extends to a representation, also denoted by $\sigma$, of the group $SU(3)$. Even when the Lie algebra representation is known, it is very difficult to determine explicitly the group representation. However the corresponding group transformation of the densities is simple.

Suppose $\Psi$ is a normalized vector in the Hilbert space that carries the unitary representation $\sigma$ of $SU(3)$. Let $\rho = M(\Psi)$ denote its corresponding density in the dual space. The group $SU(3)$ transforms $\Psi$ into $\sigma(g)\Psi$ while the density is transformed into $M(\sigma(g)\Psi)$. For the transformed density we have
\bea
\langle M(\sigma(g)\Psi)\,,Z \rangle  &=& \langle \sigma(g) \Psi \mid \sigma(Z) \mid \sigma(g) \Psi \rangle \nonumber \\
& = & \langle \Psi \mid \sigma(g)^{-1} \sigma(Z) \sigma(g) \mid \Psi \rangle \nonumber \\
&=& \langle \Psi \mid \sigma(g^{-1} Z g) \mid \Psi \rangle= \mbox{tr}\,( \rho\,g^{-1} Z g) \nonumber\\
&=& \langle g \rho g^{-1}\,,Z \rangle.
\eea
Hence the density $\rho$ transforms into $g \rho g^{-1}$, a product of three matrices. The group transformation in the dual space is called the coadjoint action and it is denoted by $\mbox{Ad}^\ast_g \rho = g \rho g^{-1}$ \cite{vogan}.

\subsection{Casimir Invariants}
The Casimir invariants, or Casimirs, are polynomials in the algebra generators that commute with all Lie algebra elements. ${\frak su}(3)$ has two independent Casimirs of quadratic and cubic orders,
\be
{\cal C}_r (\rho) = \mbox{tr}\, \rho^{\,r}, \mbox{\ for\ } r=2, 3 .
\ee
These are functions on the dual space that are invariant with respect to the coadjoint transformation, ${\cal C}_r(\mbox{Ad}^\ast_g \rho )={\cal C}_r(\rho)$

When $\rho = q-\frac{1}{2}\,i\,l$, the invariant functions are
\bea
{\cal C}_2(\rho) & = & \mbox{tr}\, q^2 - {\T\frac{1}{4}}\,\mbox{tr}\,l^2 \nonumber\\
{\cal C}_3(\rho) & = & \mbox{tr}\, q^3 - {\T\frac{3}{4}}\,\mbox{tr}\,(q\,l^2)  \label{Casimir23}.
\eea
In particular the values of the Casimirs at the diagonal density corresponding to a highest weight vector are evaluated to be
\bea
{\cal C}_2(\rho_{(\lambda \mu)}) &=& {\T \frac{2}{3}}(\lambda^2 +\lambda\mu +\mu^2) \nonumber \\
{\cal C}_3(\rho_{(\lambda \mu)}) &=& {\T \frac{1}{9}}(2\lambda^3 +3\lambda^2\mu -3\lambda\mu^2 -2\mu^3) \label{Casimirvalues}.
\eea

In quantum mechanics, observables are hermitian linear operators, while in density matrix theory, observables are real-valued functions on the dual space. The expectations of the Casimir operators with respect to a highest weight state differ from the density matrix functions ${\cal C}_r(\rho)$ by terms of lower degree. The ultimate reason for the discrepancy is that quantum fluctuations are not included in the density matrix theory. A mathematically rigorous presentation of the relationship between polynomial operators in the enveloping algebra and functions of the density matrices is given in the Appendix. 

It is important to maintain consistency within the density matrix theory and not replace the values of the Casimir functions, Eq.~(\ref{Casimirvalues}), by their quantum expectations in hopes of an improved theoretical description.  Similarly the square of the total angular momentum is $I^2$ in the density matrix theory, not $I(I+1)$.

\subsection{Admissible densities}
In the mean field approximation the admissible densities are restricted to those that lie on a level surface of the Casimirs, i.e., a surface on which the two Casimir functions are constant.
The $SU(3)$ group transformations $\mbox{Ad}^\ast_g$ leave each level surface invariant. Since any hermitian matrix $\rho\in{\frak su}(3)^\ast$ may be diagonalized by some unitary transformation $g\in SU(3)$, each level surface contains a traceless diagonal matrix. Because eigenvalues are unique, each level surface contains a unique diagonal matrix up to ordering of the real eigenvalues, which may be parameterized by nonnegative $\lambda$ and $\mu$ as in Eq.~(\ref{matrixrho}). Note that a diagonal density only corresponds to a highest weight vector when $\lambda$ and $\mu$ are also integers.

In the typical case, a level surface is six dimensional, because there are two functionally independent conditions imposed in the eight dimensional dual space. In the special cases of $\mu$ or $\lambda$ equal zero, the level surface of admissible densities is four dimensional.

\section{${\frak su}$(3) rotor states}
\label{sec:level3}
The rotation group $SO(3)$ is a subgroup of the special unitary group $SU(3)$.  A density $\rho=q-\frac{1}{2}\,i\,l$ in ${\frak su}(3)^\ast$ is transformed by a rotation $R\in SO(3)$ into the density $\mbox{Ad}^\ast_R \rho = R \rho R^{\Sc T} = R\/qR^{\Sc T}-\frac{1}{2}\,i\,R\/lR^{\Sc T}$. Since any real symmetric matrix can be diagonalized by a rotation matrix, there is a $R\in SO(3)$ such that the rotated quadrupole moment is diagonal,
\be
\bar{q}=R\/qR^{\Sc T}=\diag\,(q_1,q_2, q_3) .
\ee
The eigenvalues are unique, up to their order, which we fix to be $q_3\geq q_1\geq q_2$. From a geometrical viewpoint, $R$ rotates the laboratory frame into the body-fixed frame in which, by definition, the system's quadrupole moment $\bar{q}$ is diagonal . At the same time the laboratory angular momentum $l$ is transformed to $I=R\,\/l \, R^{\Sc T}$, which is the system's angular momentum projected onto the body-fixed principal axes. The matrix $I$ is antisymmetric, but otherwise arbitrary. In general, the angular momentum vector is not aligned with a principal axis.

The diagonal entries of $\bar{q}$ define the $(\beta, \gamma)$ deformation parameters in the body-fixed  frame \cite{bohrmott98}
\be
q_k=\beta\cos(\gamma-k\T{\frac{2\pi}{3}}),\quad k=1,2,3\,.
\ee
The chosen ordering for the eigenvalues $q_k$ corresponds to $\beta\geq 0$ and $\gamma\in [0,\pi /3]$.
The trace of any power of $q$ is a rotational scalar; the quadratic and cubic scalars simplify to
\bea
\tr(q^2)&=&{\T\frac{3}{2}}\,\beta^2\\
\tr(q^3)&=&{\T\frac{3}{4}}\,\beta^3\cos(3\gamma) ,
\eea
which are model-independent measures of deformation \cite{Cline}.

The angular momentum is a pseudovector. The vector components of the angular momentum are given by $l_i  =  \T{\frac{1}{2}}\, \ve_{ijk}\, l_{jk}$ in the laboratory frame and by $I_i  =  \T{\frac{1}{2}}\, \ve_{ijk}\, I_{jk}$ in the body-fixed frame. The rotation of the vector angular momentum $\vec{I} = R\, \vec{l}$ is equivalent to the matrix transformation $I=R\/l R^{\Sc T}$. A principal axis rotation requires that two of the three components of $\vec{I}$ are zero. A tilted rotation in a principal plane requires that one component of $\vec{I}$ is zero. But, in general, all three components of the angular momentum in the principal axis frame are nonzero. The rotational scalars of Eq.~(\ref{Casimir23}) that are quadratic in the angular momentum matrix may be expressed in terms of the vector components
\bea
\mbox{tr}\ l^2 & = & -2\, I_k I_k  \nonumber \\
\mbox{tr}\,(q\,l^2) & = & q^{}_k I_k^2 .
\eea

\subsection{Angular momenta and deformations}

The range of possible angular momenta and deformations is restricted because the admissible densities lie on a level surface of the Casimirs. Since any admissible density may be rotated to the principal axis frame, it is sufficient to solve for the admissible body-fixed densities. Such densities with total angular momentum $I$ are simultaneous solutions to the algebraic system:
\bea 
q_1+q_2+q_3&=&0 \label{trace0}\\
\sum_k\, q_k^2+{\T \frac{1}{2}}I^2 &=& {\cal C}_2 \label{casimir2}\\
\sum_k\, q_k^3-{\T \frac{3}{4}}\,\sum_k\, q^{}_k I_k^2 &=& {\cal C}_3 \label{casimir3}\\
I_1^2+I_2^2+I_3^2&=&I^2 \label{Iconstant}\,.
\eea 
This is the fundamental set of algebraic equations for ${\frak su}(3)$ density matrix theory. It is an underdetermined system of four equations for six unknowns $(q_1^{}, q_2^{}, q_3^{}, I_1^{}, I_2^{}, I_3^{})$. The fundamental system imposes the constraint that a density is admissible and has total angular momentum $I$.

The system (\ref{trace0})--(\ref{Iconstant}) determines the change in the shape of a rotating body as the angular momentum increases. Since it is a system of four equations for six unknowns $(q_k, I_k)$, analytic solutions are given uniquely only when additional assumptions are imposed. Several important special solutions can be derived including rotation about principal axes and spheroidal nuclei.

Expressed in terms of the $(\beta,\gamma)$ collective coordinates and spherical coordinates for the angular momentum in the body-fixed frame, $I_1=I\cos\phi\sin\theta$, $I_2=I\sin\phi\sin\theta$, $I_3=I\cos\theta$, $\phi\in[0,\pi]$, $\theta\in[0,\pi/2]$, the system (\ref{trace0})--(\ref{Iconstant}) is equivalent to
\bea
3\beta^2+I^2 &=& 2 {\cal C}_2 \label{c2eq}\\
3\beta^3\cos(3\gamma)+3\beta I^2 A(\phi,\theta,\gamma)\,&=& 4 {\cal C}_3,\label{c3eq}
\eea
where 
\bea
A(\phi,\theta,\gamma) & = & {\T{\frac{1}{2}}}\left( (2\sqrt{3}\sin^2\!\phi-\sqrt{3})\sin^2\!\theta\sin\gamma \right. \nonumber \\
& & \left. +(3\sin^2\!\theta-2)\cos\gamma\right). 
\eea

The deformation $\beta$ is a unique function of the angular momentum $I$. This is a kinematical property of the ${\frak su}(3)$ model which is due to the shell model compactification of the rotational algebra, i.e., the replacement of the exact quadrupole operator by the in-shell Elliott expression. The energy is irrelevant to the $\beta(I)$ functional relationship. In contrast the triaxiality $\gamma$ and the direction of the angular momentum $\vec{I}$ are not determined uniquely by the fundamental system of equations. Additional assumptions are required to derive solutions, either kinematical (rotation about principal axis or spheroidal shape) or dynamical (energy functional $E[\rho]$).

To these four equations, two more must be added to determine the quadrupole moments and angular momentum. In the following two sections, the critical points of an energy functional are used to supply the missing equations. But, in this section, many simple kinematical solutions are derived that are independent of the energy functional. Principal axis rotations require that two body-fixed components of the angular momentum vanish. Spheroidal solutions are obtained when two moments are equal. First, however, the range of the angular momentum that is compatible with the fundamental set Eqs.~(\ref{trace0})--(\ref{Iconstant}) is determined.

The function $A(\phi,\theta,\gamma)$ ranges from -1 to +1. Within the chosen interval for the angle $\gamma$, $\gamma\in[0,\pi/3]$, $\sin\gamma>0$ and $\cos\gamma>0$. The maximum values for $\sin^2\!\phi$ and $\sin^2\!\theta$ are 1, so $A_{\max}=\cos(\gamma-\frac{\pi}{3})=1$ when $\gamma=\pi/3$. Similar argument leads to $A_{\min}=-1$.

\begin{claim} The minimum angular momentum is $I=0$ and the principal axis quadrupole moments of a nonrotating body are:
\be
q_3=\frac{2\lambda+\mu}{3} \geq q_1 =\frac{-\lambda+\mu}{3} \geq
q_2=\frac{-\lambda-2\mu}{3}\,. \label{I=0ordering}
\ee \end{claim}
Note that any permutation of the three axis lengths are solutions to the system (\ref{trace0})--(\ref{Iconstant}) when $I=0$. 

The system (\ref{c2eq}, \ref{c3eq}) has a solution for the intrinsic angular momentum $I$ at both ends of the allowed intervals of values for $\phi$, $\theta$, and $\gamma$: $I=\lambda+\mu$ when $\gamma=0$, $\theta=0$, $\phi\in[0,\pi]$ and $\lambda<\mu$; $I=\lambda+\mu$ when $\gamma=\pi/3$, $\theta=\pi/2$, $\phi=\pi/2$ and $\lambda>\mu$.

\begin{proposition}\label{maximumI} The maximum allowed angular momentum is the same as the upper bound found in representation theory, $I=\lambda +\mu$.\end{proposition} 
{\bf Proof:} Reductio ad absurdum. Suppose the angular momentum can have values bigger than $\lambda+\mu$. Consider the behavior of $\beta(I)$ and $A(I)$ in (\ref{c2eq}, \ref{c3eq}) as the angular momentum increases.

For the case of oblate spheroids $\gamma=\pi/3$ and rotation around the symmetry axis ($\phi=\pi/2$, $\theta=\pi/2$) the trigonometric function $A=A_{\max}=1$. For $\lambda>\mu$ the angular momentum is $I=\lambda+\mu$, and the deformation is $3\beta=\lambda-\mu$. 
Differentiating Eq.~(\ref{c2eq}) with respect to the angular momentum $I$ and evaluating the result for $I=\lambda+\mu$ shows that as the angular momentum increases, the deformation parameter $\beta$ decreases:
\be
\frac{\mb{d}\beta}{\mb{d}I}\Bigg|_{I=\lambda+\mu}=-\frac{\lambda +\mu}{\lambda-\mu} <0
\ee
Differentiation of Eq.~(\ref{c3eq}) with respect to $I$  and evaluating the result for $I=\lambda+\mu$ leads to:
\bea
12\,\frac{(\lambda+\mu)}{(\lambda-\mu)^2}\lambda\mu &= &\frac{(\lambda-\mu)^2}{9}\,\frac{\mb{d}(\cos(3\gamma))}{\mb{d}I}\Bigg|_{\gamma=\pi/3} \nonumber \\
& & +(\lambda+\mu)^2\frac{\mb{d}A}{\mb{d}I}\Bigg|_{I=\lambda+\mu}\,.
\eea 
Assuming $\gamma(I)$ is a differentiable function, $\D{\frac{\mb{d}\gamma}{\mb{d}I}}\neq\infty$, and
\be
\frac{\mb{d}(\cos(3\gamma))}{\mb{d}I}\Bigg|_{\gamma=\pi/3}=-3\sin(3\gamma) \frac{\mb{d}\gamma}{\mb{d}I}\Bigg|_{\gamma=\pi/3} =0\,.
\ee
Therefore 
\be
(\lambda+\mu)^2\frac{\mb{d}A}{\mb{d}I}\Bigg|_{I=\lambda+\mu}=12\,\frac{(\lambda+\mu)}{(\lambda-\mu)^2}\lambda\mu >0\,,
\ee
or the function $A$ increases as the angular momentum increases beyond $I=\lambda+\mu$. But this contradicts the fact that $I=\lambda+\mu$ is a solution to the system when the function $A$ has its maximum allowed value $A_{\max}=1$. Thus, the maximum value of the angular momentum is $\lambda+\mu$.\quad\rule{2mm}{2mm}

\begin{claim} When $I=\lambda+\mu$, there is a unique solution to (\ref{trace0})--(\ref{Iconstant}) with three possible cases for the deformation: (a) $\lambda > \mu$, $\gamma = \pi/3$, $\beta=(\lambda-\mu)/3$ and noncollective oblate rotation; (b) $\lambda < \mu$, $\gamma = 0$, $\beta=(\mu-\lambda)/3$ and noncollective prolate rotation; (c) $\lambda = \mu$, $\beta=0$, $\gamma$ and the rotation axis for the sphere are undetermined. \end{claim}

From Eq.~(\ref{c2eq}) for the quadratic Casimir, the deformation $\beta$ and the total angular momentum $I$ are evidently bounded. For no rotation, $I=0$, the deformation attains its maximum $\beta_{\rm max} = 2\sqrt{\lambda^2+\lambda \mu +\mu^2}/3$. Although $\beta=0$ is a solution to the quadratic Casimir equation, it is {\em not} generally a solution to the cubic Casimir, Eq.~(\ref{c3eq}). The exception is $\lambda=\mu$.  Thus, the cubic Casimir invariant is essential to the upper bound for the angular momentum and the lower bound for the deformation. As proven above, the maximum angular momentum $I=\lambda+\mu$ corresponds to the minimum deformation $\beta_{\rm min}=|\lambda-\mu|/3$. In Fig.~\ref{fig1}, $\beta$ is plotted versus $I$; the solution curve is an ellipse.

\subsection{Analytical Solutions for Rotation Around One of the Principal Axes}

The system (\ref{trace0})--(\ref{Iconstant}) can be solved analytically for the case of general $(\lambda,\mu)$ and rotation around one of the principal axes, say the 1-axis. Assume that the angular momentum is directed along the principal $1$-axis, $I_2=I_3=0$, $I_1=I$. In this section, in contrast to the prior convention, the quadrupole moments of the principal axis solutions are not ordered. Thus the rotation axis is fixed, but $q_1$ may correspond to the short, long, or middle length axis. 

\begin{claim} For rotations about the principal $1$-axis, there are three analytical solutions of the algebraic system (\ref{trace0})--(\ref{Iconstant}):\end{claim}
\widetext
\bea
q_1&=&-\frac{\lambda+2\mu}{3},\quad q_{2,3} = \frac{\lambda+2\mu}{6}
\pm \frac{1}{2}\sqrt{\lambda^2-I^2},\quad 0\leq I \leq \lambda, \mbox{ short axis rotation}, \label{k=0band}\\
q_1&=&+\frac{2\lambda+\mu}{3},\quad q_{2,3} = -\frac{2\lambda+\mu}{6} \pm
\frac{1}{2}\sqrt{\mu^2-I^2},\quad 0\leq I \leq \mu, \mbox{ long axis rotation}, \label{k=ilevels}\\
q_1&=&-\frac{\lambda-\mu}{3}, \quad q_{2,3} = \frac{\lambda-\mu}{6} \pm
\frac{1}{2}\sqrt{(\lambda+\mu)^2-I^2},\quad 0\leq I \leq \lambda+\mu, \mbox{ middle axis rotation}. \label{middleaxisrot}
\eea
\narrowtext
Note that, since $q_1$ is the deformation along the rotation axis -- short, long, or middle -- the formulae for $q_2$ and $q_3$ can be written as
\be \label{1axissolutions}
q_{2,3}=-\frac{q_1}{2}\pm\frac{1}{2}\sqrt{I_{\max}^2-I^2}\,.
\ee
It can be seen immediately that:
\begin{enumerate}
\item The quadrupole moment along the rotation axis remains constant.
\item At the maximal $I$, viz. $I_{\max}=\lambda$, $I_{\max}=\mu$, $I_{\max}=\lambda+\mu$,
these solutions are non-collective spheroidal, i.e. axially symmetric with respect to the axis of rotation.
\item The three solutions correspond to bands rotating about the short, long, and middle axes of the $I=0$ nonrotating solution. For $\lambda>\mu$, the yrast band is described by the system rotating collectively about its short axis, $0\leq I\leq \lambda$. When the system rotates collectively about its long axis, the densities correspond to the bandheads, $0\leq I=K \leq \mu$. It is unclear if the densities describing rotation about the middle axis are found in nature; recall that such rotations are unstable in classical mechanics.
\item Triaxiality is indicated by the $\gamma$ parameter. In Fig.~\ref{fig2}, the triaxiality is plotted versus the total angular momentum for short and long axes rotation in the $(\lambda,\mu)=(8,4)$ irrep. Each band terminates in a phase transition to a noncollective rotation. For short and long axis rotations, the general expression for triaxiality is
\bea
\tan \gamma = \left\{\begin{array}{ll} \sqrt{3}\,{\D\frac{\lambda+2\mu-\sqrt{\lambda^2-I^2}}{\lambda+2\mu+3\sqrt{\lambda^2-I^2}}}, & \mbox{short} \\ [.4cm]
\sqrt{3}\,{\D\frac{\sqrt{\mu^2-I^2}}{2\lambda+\mu}}, & \mbox{long}. \end{array} \right.
\eea
\item When $\mu = 0$, ${\cal C}_2 = \frac{2}{3}\lambda^2$, ${\cal C}_3 = \frac{2}{9}\lambda^3$, $0\leq I \leq \lambda$,  and the solutions scale as $\lambda$:
\be
\frac{q_1}{\lambda}=-\frac{1}{3},\qquad \frac{q_{2,3}}{\lambda} = \frac{1}{6} \pm
\frac{1}{2}\sqrt{1-\left(\frac{I}{\lambda}\right)^2} .
\ee
When $I=0$, the nucleus is a prolate spheroid. For small values of the angular momentum, $I\ll\lambda$, the rotation is approximately collective prolate, but the shape is slightly triaxial. When the nucleus rotates with the maximum allowed angular momentum, $I_{\max}=\lambda$, it is an oblate spheroid rotating around its symmetry axis (non-collective oblate rotation). The results for the deformation as a function of the angular momentum are summarized in Table \ref{mu0}.

\item When $\lambda = 0$, ${\cal C}_2 = \frac{2}{3}\mu^2$, ${\cal C}_3 = -\frac{2}{9}\mu^3$, $0\leq I \leq \mu$, and the solutions scale as $\mu$: 
\be
\frac{q_1}{\mu}=\frac{1}{3},\qquad \frac{q_{2,3}}{\mu} = -\frac{1}{6} \pm
\frac{1}{2}\sqrt{1-\left(\frac{I}{\mu}\right)^2} .
\ee
The initial shape (at $I=0$) is an oblate spheroid. For small values of the angular momentum, $I\ll\mu$, the rotation is approximately collective oblate, but the shape is slightly triaxial. When $I$ reaches its maximum value, $I_{\max}=\mu$, the nucleus is a prolate spheroid rotating around its symmetry axis (non-collective prolate rotation). The deformation as a function of the angular momentum is given in Table \ref{lambda0}.

\item When $\lambda = \mu$, ${\cal C}_2 = 2\lambda^2$, ${\cal C}_3 = 0$. For a given angular momentum $I$, when $(q_1,q_2,q_3)$ is a solution, so is $(-q_1,-q_2,-q_3)$.
\begin{itemize}
\item $q_1 =0$ and $0\leq I \leq 2\lambda$ 
\bea
q_2 & = & \pm \frac{1}{2} \sqrt{4\lambda^2 - I^2} \nonumber \\
q_3 & = & -q_2
\eea
Here, the nucleus begins as a triaxial shape rotating around the middle axis. At $I_{\max}=2\lambda$ the nucleus turns into a sphere.
\item $q_1 = \lambda$ and $0\leq  I\leq \lambda$. 
\bea
q_2 & = & -\frac{\lambda}{2} + \frac{1}{2}\sqrt{\lambda^2 - I^2} \nonumber \\
q_3 & = & -\frac{\lambda}{2} - \frac{1}{2}\sqrt{\lambda^2 - I^2}
\eea 
In this case the nucleus is triaxial at $I=0$ and begins rotating around its long axis. At the maximum value of the angular momentum $I_{\max}=\lambda$ the nucleus is a prolate spheroid rotating non-collectively.
\end{itemize}
\end{enumerate}

\subsection{Spheroidal solutions}
Analytical solutions can be found for spheroids either rotating about a principal axis or in a principal plane. Denote the two equal quadrupole moments by $q$ and the unequal axis moment by $-2q$. The quadratic Casimir Eq.~(\ref{casimir2}) determines the deformation as a function of the angular momentum
\be
6q^2 + {1\over2}I^2 = {\cal C}_2\,. \label{spherequad}
\ee
Let $K$ denote the component of the angular momentum along the symmetry axis.

\subsubsection{Spheroids rotating around one principal axis}

\begin{claim} $K=0$: Collective rotation perpendicular to the symmetry axis.\end{claim} 
A solution must satisfy the cubic Casimir equation (\ref{casimir3})
\be
-24q^3 - 3qI^2 = 4{\cal C}_3\,.\label{spherecubick=0}
\ee
Eliminating the angular momentum $I$ from Eqs.~(\ref{spherequad}) and (\ref{spherecubick=0}) yields a single equation for the deformation, $6 q^3 - 3{\cal C}_2 q - 2{\cal C}_3 = 0.$
There is only one physical solution:
\be q=-\frac{\lambda-\mu}{3}, \qquad I^2 = 4\lambda\mu. \label{spherek=0sol}
\ee
When $\lambda > \mu$, this is a prolate spheroid; for $\lambda < \mu$, the solution is an oblate spheroid. The other two solutions to the system (\ref{spherequad}, \ref{spherecubick=0}) , $q=\frac{1}{3}\,(2\lambda+\mu)$ and $q=-\frac{1}{3}\,(\lambda+2\mu)$, are unphysical because they imply negative $I^2$.

\begin{claim} $K=I$: Noncollective rotation about the symmetry axis. \end{claim}
A solution satisfies the cubic Casimir equation (\ref{casimir3})
\be
-12q^3 + 3qI^2 = 2{\cal C}_3\,. \label{spherecubick=I}
\ee
Eliminating the angular momentum $I$ from Eqs.~(\ref{spherequad}) and (\ref{spherecubick=I}) yields a single equation for the deformation, $24 q^3 - 3{\cal C}_2 q + {\cal C}_3 = 0.$
All three solutions here are physical:
\bea
q &= & (\lambda+2\mu)/6, \quad I=\lambda ; \label{spherek=Ioblate} \\
q &= & -(2\lambda+\mu)/6, \quad I=\mu ; \label{spherek=Iprolate} \\
q &= & (\lambda-\mu)/6, \quad I = \lambda+\mu . \label{spherek=I?} 
\eea 
When $q>0$ the solution is an oblate spheroid; when $q<0$, it is prolate. Note that these noncollective spheroids are the band terminations of the principal axis solutions (\ref{k=0band}), (\ref{k=ilevels}), and (\ref{middleaxisrot}).

\subsubsection{Tilted rotation of spheroids} \label{tiltedspheroids}

\begin{claim} $K$ is between $0$ and $I$, i.e., the rotation is tilted. \end{claim}
Using the cubic Casimir equation, Eq.~(\ref{casimir3}), the squared projection $K^2$ may be solved for as a function of the angular momentum $I$,
\be
K^2 = {I^2 + 4 {\cal C}_2\over9} + {4{\cal C}_3\over9q}. \label{tiltedeq}
\ee
When $\lambda>\mu$, a prolate spheroid ($q<0$) solution exists in the interval $\mu \leq I \leq \sqrt{4\lambda\mu}$ with the squared projection of the angular momentum on the symmetry axis
\widetext
\bea
K^2 &=& \mu^2 + {1\over9}(I^2-\mu^2) + {8\over27}(\lambda-\mu)(\lambda+2\mu)
\left[ 1 -{2\lambda+\mu\over\sqrt{(2\lambda+\mu)^2-3(I^2-\mu^2)}} \right]\\
&=& \mu^2 + {\mu^2\over(2\lambda+\mu)^2} (I^2-\mu^2) -{(\lambda-\mu)(\lambda+2\mu)\over(2\lambda+\mu)^4} (I^2-\mu^2)^2 - \ldots \label{taylorprolate}
\eea
\narrowtext
For $I=\mu$, the prolate spheroid rotates noncollectively about its symmetry axis, Eq.~(\ref{spherek=Iprolate}). At a maximum $I=\sqrt{4\lambda\mu}$, the prolate spheroid is rotating perpendicularly to its symmetry axis. In Fig.~\ref{fig3} the projection $K$ is plotted versus $I$ for the $(\lambda,\mu)=(8,4)$ prolate spheroids; in Table~\ref{Mgspheroids} the projection and the deformation are given for this case. Note that the projection is approximately constant $K\approx 4$ until $I$ nears the top of the band; this is evident from the Taylor expansion (\ref{taylorprolate}). Thus there is a $K\approx\mu$ band of prolate states for $\mu\leq I < 2\sqrt{\lambda\mu}$. The band may be viewed from the top down: Start with a collective rotation of a prolate spheroid with $I=2\sqrt{\lambda\mu}$. As the body rotates more slowly, it rapidly acquires a component $K\approx4$ along the symmetry axis. As the magnitude of the angular momentum becomes smaller, its direction becomes more aligned with the symmetry axis. The band terminates in a noncollective rotation when $I=K=\mu$.

When $\lambda>\mu$ there are also tilted oblate solutions for $\lambda\leq I\leq \lambda+\mu$. The two endpoints correspond to noncollective rotation and termination of the short and middle axes bands (\ref{spherek=Ioblate}) and (\ref{spherek=I?}).

When $\lambda<\mu$, an oblate spheroid $(q>0)$ rotates with the projection
\widetext
\bea
K^2 & = & \lambda^2 + {1\over9}(I^2-\lambda^2) - {8\over27}(\lambda-\mu)(\mu+2\lambda)
 \left[ 1 -  {2\mu+\lambda\over\sqrt{(2\mu+\lambda)^2-3(I^2-\lambda^2)}} \right]  \\
& = & \lambda^2 + {\lambda^2\over(2\mu+\lambda)^2} (I^2-\lambda^2) +{(\lambda-\mu)(\mu+2\lambda)\over(2\mu+\lambda)^4} (I^2-\lambda^2)^2 - \ldots  \label{tayloroblate}
\eea
\narrowtext
For $I=\lambda$, the solution is an oblate spheroid rotating noncollectively about its symmetry axis. At a maximum $I=\sqrt{4\lambda\mu}$, the oblate spheroid is rotating perpendicularly to its symmetry axis. When $\lambda<I\ll\sqrt{4\lambda\mu}$, $K$ is approximately $\lambda$. From the Taylor expansion (\ref{tayloroblate}), the projection is approximately constant $K\approx\lambda$ until one nears the top of the band.

If $\lambda=\mu$, the cubic Casimir is zero, and
\be
K^2 = \mu^2 + {I^2-\mu^2\over9} .
\ee
As the angular momentum varies from a minimum $I=\mu$ to a maximum $I=2\mu$, the projection of the angular momentum on the symmetry axis varies from $K=\mu$ to $K={4\over3}\mu$.

\section{Energy functional I} \label{energyfunctional1}
The particular solutions enumerated in \S~\ref{sec:level3} correspond to simple kinematical situations. The actual physical densities must be determined from a dynamical argument. An equilibrium density for a rotating body in the ${\frak su}(3)$ model is a critical point of the energy functional $E[\rho]$ on the surface of admissible densities. For a body with constant moments of inertia, the energy is
\be
E[\rho] = A_1 I_1^2 + A_2 I_2^2 + A_3 I_3^2 , \label{energyrho1}
\ee
where $A_1, A_2, A_3$ are real constants. Suppose the total angular momentum is $I$, the admissible densities lie on the surface determined by the quadratic and cubic Casimirs ${\cal C}_k$, and $\rho$ is a critical point of $E[\rho]$ on the surface. By the Lagrange multiplier theorem, there are four real constants $a, b, c, d$ associated with each of the constraint equations (\ref{trace0}--\ref{Iconstant}) such that $\rho$ is a critical point of the functional
\bea
F[\rho] & = & E[\rho] + a \left(q_1+q_2+q_3\right) + b \left(\sum_k q_k^2+{\T \frac{1}{2}}I^2 - {\cal C}_2\right) \nonumber \\
& & + c \left(\sum_k q_k^3-{\T \frac{3}{4}}\, \sum_k q_k I_k^2 -  {\cal C}_3\right) \nonumber \\
& & + d \left(I_1^2+I_2^2+I_3^2 - I^2\right) \label{constrainedenergy}
\eea
with a free variation on $(q_1, q_2, q_3, I_1, I_2, I_3)$. Altogether there are 10 unknowns, including the Lagrange multipliers, that must satisfy the system of 10 equations,
\bea
0 &=& {\partial F \over \partial q_k} = a + 2 b q_k^{} + c\left( 3q_k^{2} - {3\over4}I_k^2 \right) \label{sixeqns1} \\
0 &=& {\partial F \over \partial I_k} = 2 I_k^{}\left( A_k^{} -{3\over4} c q_k^{} +d \right) \label{sixeqns2}
\eea
plus the four constraint equations. Although the number of equations in the system is large, each is just a polynomial of low degree in the variables. Thus analytic solutions may be determined rather easily in many cases.

Indeed each of the kinematical solutions enumerated in \S~\ref{sec:level3} are particular solutions to the Lagrange multiplier system:
\begin{enumerate}
\item Rotations about a principal axis, say the $1$-axis, given by Eqs.~(\ref{k=0band}, \ref{k=ilevels}, \ref{middleaxisrot}) are critical points of (\ref{constrainedenergy}) when $d=A_1$ and $a=b=c=0$.
\item Spheroidal solutions, $q_2=q_3=q\neq0$ and $q_1=-2 q$, are obtained when $a=b=c=0$ and
\be 
I_1 (A_1-d) = I_2 (A_2-d) = I_3 (A_3-d) = 0. \label{classicalrotor}
\ee
Principal axis rotation is one type of solution. When two energy parameters are equal, say $A_1=A_3$, tilted rotations are also critical points for $d=A_1$ and $I_2=0$.  
\end{enumerate}

It is interesting to compare the classical rigid rotor to an ${\frak su}(3)$ rotor. A classical rigid rotor in equilibrium is a critical point of the energy (\ref{energyrho1}) subject to the constraint $\sum I_k^2 = I^2$. This is mathematically equivalent  to the ${\frak su}(3)$ constraint problem when $a=b=c=0$, viz., Eq.~(\ref{classicalrotor}). The ${\frak su}(3)$ problem allows for additional critical points. The mathematical origin of the extra solutions is the different constraint system for the admissible ${\frak su}(3)$ surfaces. The physical origin is related ultimately to the shell model, which is responsible for the ${\frak su}(3)$ quadrupole operator (\ref{quaddef}) instead of the major shell-mixing quadrupole operator $\sum r_\alpha^2 Y^{(2)}_\mu(\Omega_\alpha)$.

Suppose the kinematical solutions corresponding to $a=b=c=0$ are excluded from consideration. In this case a solution to Eq.~(\ref{sixeqns1}) requires
\bea
0 &=& 4\,\left(q_2-q_1\right)\,\left(q_3-q_1\right)
 \,\left(q_3-q_2\right) \\
& & + I_1^2\,(q_2-q_3)
+ I_2^2\,(q_3-q_1) + I_3^2\,(q_1-q_2) . \nonumber
\eea 
In conjunction with Eqs.~(\ref{casimir3}, \ref{Iconstant}) this condition enables the determination of the body-fixed angular momentum components in terms of the quadrupole moments:
\be
I_1^{2}=\frac {I^{2}}{3} + \frac {4\,\left(q_3^{4}+2\,q_2\,q_3
 ^{3}+2\,q_2^{3}\,q_3+q_2^{4}\right)
 -2\,q_1{\cal C}_3}{3\,\left(q_3^{2}+q_2\,q_3
 +q_2^{2}\right)} \label{genangmom}
\ee
and similarly for $I_2, I_3$ by cyclic permutation of $1, 2, 3$.
After eliminating the Lagrange multipliers the system reduces to a single independent equation,
\be
0 = I_2 \left( (A_2-A_3) q_1 + (A_3-A_1) q_2 + (A_1-A_2) q_3  \right) ,
\ee
where $I_1, I_3$ are assumed nonzero.
There are two cases to consider as one of the two factors in the above equation must vanish. 

\begin{claim}
Principal plane rotations of triaxial bodies.
One of the components of the angular momentum vanishes, say the $2-$axis projection,
$I_2 = 0$.
Every critical point of the principal plane system is determined by a real root $q_1$ of the eighth degree polynomial
\widetext
\bea
 0 & = & 576\,q_1^{8}+288\,\left(I^{2}-2\,{\cal C}_2\right)\,q_1
 ^{6}+192\,{\cal C}_3\,q_1^{5}+\left(36\,I^{4}-168\,{\cal C}_2
 \,I^{2}+192\,{\cal C}_2^{2}\right)\,q_1^{4} +\left(72\,{\cal C}_3
 \,I^{2}-144\,{\cal C}_2\,{\cal C}_3\right)\,q_1^{3} \nonumber \\
& & +\left(-12
 \,{\cal C}_2\,I^{4}+48\,{\cal C}_2^{2}\,I^{2}+16\,{\cal C}_3^{2}-48
 \,{\cal C}_2^{3}\right)\,q_1^{2}+\left(8\,{\cal C}_2^{2}\,{\cal 
C}_3-4\,{\cal C}_2\,{\cal C}_3\,I^{2}\right)\,q_1 +{\cal C}_2^{
 2}\,I^{4}+\left(4\,{\cal C}_3^{2}-4\,{\cal C}_2^{3}\right)\,I^{2}\nonumber \\
& &-8
 \,{\cal C}_2\,{\cal C}_3^{2}+4\,{\cal C}_2^{4} . 
\eea
When $q_1$ is a real root, the other two quadrupole moments and angular momentum components are given by 
\bea
q_2 & = & \frac {\left(I^{2}-2\,{\cal C}_2\right)\,\left(6\,q_1^{2}
 -{\cal C}_2\right)}{4\,\left(6\,q_1^{3}+{\cal C}_3\right)}, \qquad q_3  =  -q_2-q_1 \\
 I_3^2 & = & \frac {\left(\,q_2-\,q_3\right)\,I^{2} - 4
\left[\, q_1^{2}\,\left(\,q_2^{}-\,q_3^{}\right)+q_2^{2}\,\left(\,q^{}_3-\,q^{}_1\right)+q_3^{2}\,\left(\,q_1^{}-\,q^{}_2\right) \,\right] }{3\,q_2}, \qquad  I_1^2 =  I^2 - I_3^2 .
\eea
\narrowtext
A valid solution is attained whenever the squares of the angular momentum components $I_1^2, I_3^2$ are nonnegative.
\end{claim}

Since $I_2=0$ there is a reflection symmetry: when $q_1$ is a valid solution, $q_3$ also satisfies the eighth degree polynomial. Thus, solutions come in pairs except in the spheroidal case $q_1=q_3$. To eliminate this trivial redundancy, we choose $q_1\leq q_3$. In Table \ref{tiltedtriaxialband} two sequences of solutions are given for the case $(\lambda,\mu)=(8,4)$. There are no valid solutions for angular momentum $I$ less than $\mu=4$. When $I=4$ the noncollective prolate solution (\ref{spherek=Iprolate}) begins a  band of principal plane triaxial solutions that terminates at $I\approx 10.08$. When $I=8$ a second band emerges starting from the noncollective oblate state (\ref{spherek=Ioblate}) and also terminating at $I\approx 10.08$. Above the critical angular momentum $I\approx 10.08$, the pair of real roots turns into a pair of complex conjugate roots.

There is a third band of principal plane triaxial rotors that begins at $I=\sqrt{4\lambda\mu}$ which is the termination of the sequence of prolate spheroidal solutions (\ref{spherek=0sol}). This band terminates at $I=\lambda+\mu$ in the noncollective oblate state. In Figure \ref{fig4} the projection $I_1$ of the angular momentum on the short axis for this band is plotted versus the angular momentum $I$ for the case $(\lambda,\mu)=(8,4)$. In Figure \ref{fig3} this band is shown as the continuation of the prolate spheroid band to triaxial principal plane rotors when $I>\sqrt{4\lambda\mu}$; $I_2=K=0$ is the projection of the angular momentum on the long axis.

\begin{claim}
Constant-$\gamma$ bands.
If all three components of the angular momentum are nonzero, $I_k\neq0$, then 
\be
(A_2-A_3) q_1 + (A_3-A_1) q_2 + (A_1-A_2) q_3 = 0 .
\ee
When two of the energy coefficients are equal, say $A_2=A_3$, this equation forces a spheroidal solution, $q_2=q_3$. When the coefficients are all distinct, the system may be solved for the quadrupole moments: For $i, j, k$ cyclic,
\be
q_i  =  (2\,A_i - A_j - A_k) u,
\ee
\be
u^2 = {2\,{\cal C}_2 - I^2\over 12\,\left( A_1^2+  A_2^2+ A_3^2 - A_1\,A_2 - A_2\,A_3 - A_3\,A_1  \right)} .
\ee
\end{claim}
 
Since the ratios among the quadrupole moments are constant, the triaxiality parameter $\gamma$ is fixed and independent of the angular momentum. The angular momentum components in the intrinsic frame are determined from Eq.~(\ref{genangmom}); a valid solution demands that the square of each angular momentum component is nonnegative. These fixed-$\gamma$ bands typically start at rather large angular momentum and terminate at the maximum angular momentum $I=\lambda+\mu$.

\section{Energy functional II}

A second natural choice for the energy functional is a polynomial in the ${\frak so}(3)$ integrity basis of ${\frak su}(3)$. By definition, any rotationally invariant polynomial in the ${\frak su}(3)$ enveloping algebra is a function of the ${\frak so}(3)$ integrity basis. The basis consists of five functionally-independent polynomials: the two Casimirs, ${\cal C}_2, {\cal C}_3$, the square of the total angular momentum $L^2$, and two scalars of degrees three and four \cite{judd74},
\bea
X_3 &=&L_i {\cal Q}^{\Sc(2)}_{ij} L_j \\
X_4 &=&L_i {\cal Q}^{\Sc(2)}_{ij}{\cal Q}^{\Sc(2)}_{jk} L_k \, .
\eea
The eigenvalues of the hermitian operators $X_3$ and $X_4$ in a representation are not integers \cite{racah}. On a level surface the Casimirs are constants and the other three integrity basis members may be evaluated in the principal axis frame, $L^2 = I^2$, $X_3 = q_k^{} I_k^2$, $X_4 = q_k^2 I_k^2$. A simple energy functional used in nuclear structure applications is a linear combination of $I^2$, $X_3$, and $X_4$,
\be
E[\rho] = A I^2 + B X_3 + C X_4 ,
\ee 
where $A, B, C$ are real constants \cite{ros84}. Note that the moments of inertia for this energy functional are not constant, and they  depend on the deformation. When $C\neq0$, the energy may be expressed conveniently as
\be \label{energy2}
E[\rho] = A I^2 + C (\nu X_3 + X_4),
\ee
where $\nu = B/C$.  Since $I^2$ is constant on the constraint surface and $C$ determines an energy scaling, the critical points depend only on the real parameter $\nu$.  The critical points of  (\ref{energy2}) on the constraint hypersurface (\ref{trace0} -- \ref{Iconstant}) in the dual space are determined by the Lagrange multiplier theorem in a way similar to Eq.~(\ref{constrainedenergy}). 

Once again, the kinematical solutions of \S~\ref{sec:level3} are critical points of (\ref{energy2}).
\begin{enumerate}
\item Rotations about a principal axis, viz., Eqs.~(\ref{k=0band}, \ref{k=ilevels}, \ref{middleaxisrot}), are critical points for the multipliers, $a = 3 c q_2 q_3$, $b = 3 c q_1/2$, $d = q_1 (q_1 + \nu +3 c/4)$, and 
\be
c = \frac{I^2 (2 q_1 + \nu)}{3\left( 2q_1^2+ q^{}_2 q^{}_3 - I^2/4\right)} .
\ee
When the denominator in the above expression for $c$ is zero, there is still a principal axis solution, but only when the parameter $\nu = -2 q_1$. For example, the band given by Eq.~(\ref{k=0band}) for $0\leq I\leq \lambda$ yields
\be
c = -{I^2 (2\lambda+4\mu -3 \nu)\over 9\mu(\lambda+\mu)}.
\ee
When $\mu =0$ the energy parameter must be $\nu = 2\lambda/3$.   \\

\item Spheroidal solutions are critical points only when $I_2 = I_3$. In this case the Lagrange multipliers may be chosen to yield the spheroidal solutions.
\end{enumerate}

The equations for general triaxial rotors are obtained after eliminating the Lagrange multipliers from the system. Ignoring the special case of spheroidal solutions, a solution requires that the rotation is in a principal plane.

\begin{claim}
Principal plane rotations. When $I_2=0$ the body-fixed angular momentum components $I_1, I_3$ are determined from Eqs.~(\ref{casimir3}, \ref{Iconstant}). In addition to the constraint equations (\ref{trace0}, \ref{casimir2}), the quadrupole moments satisfy
\bea
0 & = & 12(q_1^3+q_3^3 - 4q^{}_1q^{}_2q^{}_3) - 3I^2q^{}_2 \nonumber \\
& & +6\nu (q^{}_3-q^{}_2)(q^{}_1-q^{}_2) +2{\cal C}_3 .
\eea
\end{claim}
Critical points of the energy functional are shown in Figures~\ref{fig5}, \ref{fig6}, \ref{fig7} for $(\lambda, \mu) = (8,4)$ and for three values of $\nu$:  $20/3$, $0$, and $-4/3$. For each chosen $\nu$ there are two plots in the figure: one of the triaxiality parameter $\gamma$ versus the angular momentum, and the other of the square of the ratio $I_1 / I$ versus the angular momentum.

In these figures only the principal plane Solutions V.1 are drawn. There are, of course, the usual kinematical bands for principal axis rotations and spheroids which are not shown.  The equations are symmetrical under a simultaneous interchange of $q_1^{}$ and $q_3^{}$,  $I_1^{}$ and  $I_3^{}$.  As a result, the plots of $(I_1/I)^2$ are
reflection-symmetric with respect to the $(I_1/I)^2 = 0.5$ horizontal line. One of the curves may be regarded as a plot for $(I_1/I)^2$ while its reflection is a plot of $(I_3/I)^2$. When $(I_1/I)^2 = 0.5$ the tilting of the rotation axis in the $1-3$ plane is a maximum; when $I_1/I = 0$ or $1$, the rotation is about a principal axis. 

If $\nu = 20/3$, there are two bands, Figure~\ref{fig5}. The first band begins at the triaxial $I=0$ state and ends at $I=8$ as a noncollective oblate spheroid. Note that maximal tilting is attained for the $I=0$ density -- in vivid contrast to the principal axis solution. The second band starts at $I=4$ as a noncollective prolate spheroid and ends at $I=12$ as a noncollective oblate spheroid. Although the two end points of this band are principal axis rotations, the densities in between are tilted as the direction of the angular momentum vector varies continuously from alignment with the long axis ($I=4$) to alignment with the short axis ($I=12$).

When $\nu=-4/3$ there are two bands, but there is a gap between them. The first band begins as a triaxial
maximally-tilted rotor at $I=0$ and ends as a noncollective prolate spheroid at $I=4$. There is a sharp phase transition to the $I=4$ principal axis rotor from a highly tilted configuration. The second band is oblate-like from $I=8$ to $I=12$ while the rotation is slightly tilted.

If $\nu=0$ there are two bands, no solutions below $I=4$, and a gap between the bands. The first band begins as a triaxial, maximally-tilted rotor at $I=4$, increases in angular momentum to $I\approx 6$ before falling back to the principal axis noncollective prolate solution at $I=4$. The second band begins at $I=8$ and ends at $I=12$; these oblate-like state are only partially tilted.

\section{Comparison with representation theory}

The correspondence between density matrix theory and the irreducible representations of ${\frak su}(3)$ is quite close, but not perfect. In this section the similarities and differences between the two are discussed. 

Among the continuous family of surfaces of admissible densities, a representation may be associated only with those that satisfy a generalized Bohr-Sommerfeld quantization rule \cite{Souriaufr}. Consider the diagonal density matrix $\rho_{(\lambda \mu)}$, Eq.~(\ref{matrixrho}), which is contained uniquely in each surface. For each group element in the subgroup of diagonal matrices
\be
g = \mbox{diag}\left( e^{i\theta_1}, e^{i\theta_2}, e^{i\theta_3}  \right) ,\ \det g = 1 ,
\ee
where $\theta_1, \theta_2, \theta_3$ are real numbers, define the complex number
\bea
\chi^{(\lambda, \mu)}(g) & = & \exp\left\{ i(2\lambda+\mu)\theta_1/3 + i(-\lambda+\mu)\theta_2/3 \right. \nonumber \\
& & \left. - i(\lambda+2\mu)\theta_3/3\right\} .
\eea
The quantization condition is that $\chi^{(\lambda, \mu)}(g)$ is a character of the diagonal subgroup, i.e., whenever $g$ is the identity element of the group, $\chi^{(\lambda, \mu)}(g)$ must equal one. When $g = \mbox{diag}(e^{i\theta}, e^{-i\theta}, 1)$ is a closed circle, $0\leq\theta\leq2\pi$, the character $\chi^{(\lambda, \mu)}(g) = e^{i\lambda\theta}$ must equal one at $\theta=2\pi$, or $\lambda$ is an integer. Similar reasoning for $g = \mbox{diag}(1,e^{i\theta}, e^{-i\theta})$ shows that $\mu$ must be an integer.  A level surface of the Casimir invariants satisfies the Bohr-Sommerfeld quantization rules if and only if $\lambda$ and $\mu$ are nonnegative integers. Hence the surfaces of admissible densities that satisfy the quantization condition are in one-to-one correspondence with the irreps.

The irreducible representations of the compact algebra ${\frak su}(3)$ are finite dimensional; hence, the spectrum of any hermitian operator acting on the representation space is bounded. On a surface of admissible densities, the range of any real-valued continuous function is bounded from above and below, because these surfaces are closed and bounded manifolds. Thus the values of physical ${\frak su}(3)$ observables have a bounded range, whether hermitian operators acting on irreducible representation spaces or real-valued functions on surfaces of admissible densities. In particular, the range of the total angular momentum is the same bounded interval, $0 \leq I \leq \lambda + \mu$, in an irrep of ${\frak su}(3)$ and on the corresponding level surface of the Casimirs in the dual space  ${\frak su}(3)^\ast$. 

The differences between representation theory and density matrix theory are more evident from a comparative analysis of the angular momentum decompositions. For both ${\frak su}(3)$ irreducible representations and the adiabatic rotational model, the decomposition is achieved by angular momentum projection from a fixed intrinsic state. When the intrinsic state has a sharp quadrupole deformation, as it does in the adiabatic rotational model, the projected states $| I K M \rangle$ are orthogonal. In addition, each projected state is an eigenstate, belonging to the eigenvalue $K^2$, of the hermitian operator $\hat{I}_3^2$ that corresponds to the square of the third component of the body-fixed angular momentum \cite{rosrowe77},
\be
(q^{}_3-q^{}_1)(q^{}_3-q^{}_2) \hat{I}_3^2=\hat{X}_4 + q_3 \hat{X}_3 + q_1 q_2 I^2, \label{ksqrd}
\ee
where the quadrupole moments $(q^{}_1, q^{}_2, q^{}_3)$ are real constants determined by the intrinsic state.

In contrast, angular momentum projection from a ${\frak su}(3)$ highest weight vector yields a nonorthogonal basis that requires Gram-Schmidt orthogonalization \cite{Vergados}. Moreover $K^2$ is not an eigenvalue of $\hat{I}_3^2$ which, indeed, is no longer well-defined for ${\frak su}(3)$. These difficulties arise because the ${\frak su}(3)$ intrinsic state, the highest weight vector in an irrep, does not have sharp values for the quadrupole moments. However, when a vector's angular momentum $I$ is small compared to $\lambda$ and/or $\mu$, its deformation is little changed from that of the highest weight vector, Eq.~(\ref{casimir2}). Thus, for $I\ll \max(\lambda, \mu)$, the ${\frak su}(3)$ projected states are approximately orthogonal and $K^2$ is an approximately good quantum number.
According to Elliott \cite{Elliott}, the possible values of $K$ in an ${\frak su}(3)$ irrep $(\lambda,\mu)$ are
\be
K=\mb{min}(\lambda,\mu), \mb{min}(\lambda,\mu)-2,...,1\;\mb{or}\;0. \label{Elliottkbands}
\ee
The sequence of angular momentum states associated with a $K$-band is:
\be \label{elliottrules}
\ba{rcl}
K=0&:&I=\mb{max}(\lambda,\mu), \mb{max}(\lambda,\mu)-2,...,1\;\mb{or}\;0\\
K\neq 0&:&I=K, K+1,...,K+\mb{max}(\lambda,\mu) .
\ea
\ee

The density matrix theory shares some properties with both the geometrical rotational model and ${\frak su}(3)$ irreducible representations. The densities have well-defined values for the deformation, but, on a level surface of the ${\frak su}(3)$ Casimirs, the deformation depends on the angular momentum. In ${\frak su}(3)$ density matrix theory, Eq.~(\ref{ksqrd}) for $I_3^2$ is an identity when the operators are replaced by their corresponding functions on the dual space and, instead of remaining constant, the quadrupole moments now vary. For small angular momentum, as long as the deformation is constant, the geometrical model, the density matrix theory, and the irreducible representations of ${\frak su}(3)$ are equivalent theoretical descriptions. For larger angular momentum, the connection between the ${\frak su}(3)$ and geometrical models breaks down.  Yet, because the deformation changes on a surface of admissible densities, the matrix theory maintains its correspondence with ${\frak su}(3)$ irreps.

Suppose $\lambda>\mu$. In the density matrix theory, the band heads $0\leq I=K \leq \mu$ are densities rotating about the long principal axis, Eq.~(\ref{k=ilevels}); the last bandhead is at $I=K=\mu$ when the body is a prolate spheroid rotating about its symmetry axis. The band heads correspond to the Elliott labeling of $K$ bands, Eq.~(\ref{Elliottkbands})

The Elliott $K=0$ band corresponds to the density matrices which describe bodies rotating about the short principal axis, Eq.~(\ref{k=0band}); this band terminates at $I=\lambda$ when an oblate spheroid rotates about its symmetry axis, Eq.~(\ref{elliottrules}).

For $K\neq0$ bands, the relationship between density matrix and representation theory is not as clear. An obvious candidate for the Elliott $K=\mu$ band is the sequence of prolate spheroid densities which attains a maximum $I=\sqrt{4\lambda\mu}$, \S\ \ref{tiltedspheroids}. At this maximal $I$ the body rotates collectively about its short axis. Because the projection of the angular momentum on the symmetry axis does not maintain a constant value $K=\mu$, Figure~\ref{fig3}, the geometrical interpretation as a fixed $K$ band is not supported. In addition the density matrix band terminates before $I= \lambda+\mu$. In \S~\ref{energyfunctional1} a sequence of triaxial principal plane rotational densities which starts at $I=\sqrt{4\lambda\mu}$ and terminates at $I=\lambda+\mu$ was found. In our view the density matrix theory indicates that the Elliott $K=\mu$ band in a ${\frak su}(3)$ irrep does not exist in the strict geometrical model sense, although it is a useful concept when $\mu \leq I < \sqrt{4\lambda\mu}$. 

For the other $K$ bands, there are more discrepancies. In particular, $K$ bands in the density matrix theory need not terminate at $I=K+\lambda$. However, when $\lambda$ is large compared to both $\mu$ and the angular momentum $I$, approximate analytic solutions to the fundamental system, Eqs.~(\ref{trace0}--\ref{Iconstant}), can be found that describe tilted rotation of  triaxial rotors. For $\lambda \gg I > K$, there are approximate solutions for which the quadrupole moments in the principal axis frame are
\bea
q_1 & = & \frac {2\,\lambda+\mu}{3}-\frac {I^{2}-K^{2}}{2\,\left(2\,\lambda +\mu\right)} \nonumber \\
q_2 & = &  -\frac {2\,\lambda+\mu}{6}+\frac {\sqrt{\mu^{2}-K^{2}}}{2}+\frac {I^{2}-K^{2}}{4\,\left(2\,\lambda +\mu\right)} \\
q_3 & = &  -\frac {2\,\lambda+\mu}{6}-\frac {\sqrt{\mu^{2}-K^{2}}}{2}+\frac {I^{2}-K^{2}}{4\,\left(2\,\lambda +\mu\right)} \nonumber
\eea
and the body-fixed projections of the angular momentum are
\bea
I_1^2 &=&  K^{2} + {\mu^{2}\over\left(2\,\lambda+\mu\right)^{2}} \left(I^{2}-K^{2}\right) \nonumber \\
I_2^2 &=&  {1\over2} \left(I^{2}-K^{2}\right)\,\left(1-\frac {\mu^{2}}{\left(2\,\lambda+\mu\right)^{2}}\right) \\
I_3^2 &=&  {1\over2} \left(I^{2}-K^{2}\right)\,\left(1-\frac {\mu^{2}}{\left(2\,\lambda+\mu\right)^{2}}\right) . \nonumber
\eea
The quadratic Casimir equation is satisfied to o($\lambda^{-2}$) and the cubic Casimir equation to $o(\lambda^{-3})$.
For $\lambda$ large compared to $I$ and $\mu$, these triaxial rotors are approximately prolate spheroids. The solutions form a band in which the component of the angular momentum along the long near-symmetry axis is approximately $K$.

To distinguish among basis vectors with the same total angular momentum in an ${\frak su}(3)$ irrep, states may be chosen to be simultaneously eigenstates of some element of the ${\frak su}(3)\supset{\frak so}(3)$ integrity basis, e.g., $X_3$, $X_4$ or some more complicated function of the integrity basis \cite{judd74}. Diagonalizing just $X_3$ is the simplest choice \cite{Moshinsky}. For axially symmetric states the hermitian operator $\hat{I}_3^2$ is related to  $X_3$ \cite{rosrowe77},
\be
\hat{I}_3^2 = \frac{1}{3} ( \frac{1}{2}\,\det Q^{\Sc(2)})^{-1/3} X_3 +  \frac{1}{3} I^2\,.
\ee

A stringent test for the density matrix method is to compare the eigenvalues of $X_3$ calculated in representation theory with the simple geometrical densities. In Table \ref{Mg24K0X3}, the smallest eigenvalues of $X_3$ for each angular momentum state from $I=0$ to $I=8$ in the $(8,4)$ irrep are compared to the $X_3$ values for $K=0$ geometrical densities (\ref{k=0band}). Since $X_3$ is a cubic polynomial in the enveloping algebra and not an element of the algebra itself, the evaluation $\varphi(X_3)$ of this function in the dual space must be defined now. These values are calculated for rotation about the short principal $1$-axis by
\be
\varphi(X_3) = q^{}_1 I^2 . \label{prinaxisx3}
\ee
The error in the density matrix calculation rises to a maximum of just 10\% for the $I=8$ state.

For the $I=12$ state of the $(8,4)$ irrep, the eigenvalue of $X_3$ is $-234$; for the noncollective rotation of the oblate spheroid density, $\varphi(X_3)= (\mu-\lambda)(\lambda+\mu)^2 /3 = -192$.

For the bandheads $I=K=0, 2 ,4$, the states in the representation space are taken to be the eigenvectors belonging to the maximal eigenvalues of $X_3$, $0, 25, 108$, respectively. For the corresponding geometrical densities given by rotation about the long principal axis, $\varphi(X_3) = 0, 27, 107$, as calculated via the analytic formula (\ref{k=ilevels}). The error is negligible for small angular momentum.

In the representation space the ``$K=4$" band is chosen to be the sequence of angular momentum states belonging to the maximal eigenvalue of $X_3$ for each angular momentum. In Table \ref{Mg24K4X3}, these eigenvalues are compared to the values of $\varphi(X_3)$ for the sequence of tilted prolate spheroids, 
\be
\varphi(X_3) = q \left( I^2 - 3 K^2 \right),
\ee
where $q$ is the moment for the equal short axes and $K$ is the component of the angular momentum along the long symmetry axis.

The ``$K=2$" band in the representation space is taken as the sequence of states with intermediate values for the $X_3$ eigenvalue, see Table \ref{Mg24K2X3}. In this case the geometrical densities are assumed to be principal plane triaxial rotors, $I_2=0$ and $I_3=K=2$ where the $3$-axis is the long axis. The complex values at the end of the $K=2$ band for the values of the quadrupole matrix are due to the restriction of the rotation in one plane. They show that the assumption $I_2=0$ is not correct. There are non-zero projections of the angular momentum on all three intrinsic axes.

When $\mu = 0$, e.g., $^{20}$Ne for which $(8,0)$ is the dominant irrep, the angular momentum is multiplicity-free, $I = 0, 2, \dots, \lambda$. In this case, Bargmann and Moshinsky \cite{Bargmann1,Bargmann2} have given an analytic formula for the $\hat{X}_3$ eigenvalues: $X_3 = (\lambda/3 + 1/2) I (I+1)$. The density matrix approximation is $X_3 = q_1 I^2 = (\lambda/3) I^2$. The difference is due to the omission of commutator terms of lower degree in the density matrix approximation -- $I^2$ instead of the quantum $I(I+1)$, and $0$ instead of $1/2$.

\section{Comparison with cranked anisotropic oscillator}

In conventional mean field theory nuclear rotational motion is modeled in a simple way by cranking the anisotropic harmonic oscillator around one axis. For rotation with constant angular velocity $\omega$ around the 1-axis, the  Hamiltonian (Routhian) for one nucleon with mass $m$ in the rotating frame is 
\be \label{routhian1}
h^{\omega}=-\frac{\hbar^2}{2m}\Delta +\frac{1}{2}m(\omega_1^2 x_1^2+\omega_2^2 x_2^2 +\omega_3^2 x_3^2)-\omega \hat{I}_1
\ee
where $\omega_k$ $(k=1,2,3)$ are the oscillator frequencies. For many fermions the model wavefunction is a Slater determinant given by occupying the orbitals of $h^{\omega}$, and it is an eigenstate of the one-body operator $H^{\omega} = \sum_\alpha h^{\omega}$, where the sum is over the particles. When the deformation is not too large, mixing between major oscillator shells can be ignored \cite{bohrmott98,szymanski}, and, as a function of the total angular momentum $I$, the energy of a system of many nucleons in the anisotropic potential simplifies to 
\bea
E(I) & = & \hbar \omega_1 \Sigma_1 +  +  \frac{1}{2} \hbar \omega_2  \left(\Sigma_2 +  \Sigma_3 - \sqrt{I_{\rm max}^2 - I^2 } \right)  \nonumber \\
& & +  \frac{1}{2} \hbar \omega_3  \left(\Sigma_2 +  \Sigma_3 + \sqrt{I_{\rm max}^2 - I^2 } \right) ,
\eea
where  $\Sigma_k$ denotes the sum of the quanta $(n_k+1/2)$ over all occupied orbitals. $I_{\rm max} = | \Sigma_2 - \Sigma_3 |$ is the maximum angular momentum of the rotational band, $0\leq I \leq I_{\rm max}$.
Applying Feynman's lemma,
\be \label{feynman}
\left\langle\frac{\partial H^\omega}{\partial\omega_k}\right\rangle=\frac{\partial E}{\partial\omega_k}\,,
\ee
the expectations of the  dimensionless quadrupole moment in the rotating frame are  
\bea
\langle q_1\rangle&=&\frac{1}{3} \left(2\Sigma_1-\Sigma_2-\Sigma_3\right) \\
\langle q_2\rangle&=&\frac{1}{3} \left(-\frac{1}{2}(2\Sigma_1-\Sigma_2-\Sigma_3)-\frac{3}{2}\sqrt{I_{\rm max}^2-I_1^2}\right) \\
\langle q_3\rangle&=&\frac{1}{3} \left(-\frac{1}{2}(2\Sigma_1-\Sigma_2-\Sigma_3)+\frac{3}{2}\sqrt{I_{\rm max}^2-I_1^2}\right) .
\eea
These expectations are exactly the values of the quadrupole moment found with the density matrix method, Eqs.~(\ref{k=0band}, \ref{k=ilevels}, \ref{middleaxisrot}). For example, when $\Sigma_1 \geq \Sigma_2 \geq \Sigma_3$, the rotation is about the long axis, Eq.~(\ref{k=ilevels}), where $\lambda=\Sigma_1-\Sigma_2$, $\mu=\Sigma_2-\Sigma_3 = I_{\max}$. But their derivations involve different assumptions. The ${\frak su}(3)$ density derivation shows that principal axis solutions are an immediate consequence of kinematics and the restriction to ${\frak su}(3)$ admissible densities. The constraints imposed by the quadratic and the cubic Casimirs are essential. The principal axis solutions are critical points for both energy functionals I and II. The natural inference is that the principal axis solutions should be critical points for any physically reasonable energy functional. In the density matrix theory the energy is proportional to $I^2$ for both functional I and II,
\be
E_I(I) = q_1 I^2, \qquad E_{II}(I) = \left( A+ B q_1 + C q_1^2 \right) I^2 .
\ee

In contrast the anisotropic oscillator derivation is based on a specific assumption about the energy.  When major shell mixing is ignored, the cranked anisotropic oscillator is compatible with ${\frak su}(3)$ dynamical symmetry. By making a canonical (symplectic group) transformation and expressing the ${\frak su}(3)$ generators in the transformed basis, the Hamiltonian of the cranked anisotropic oscillator (CAO) becomes an element of the algebra. Hence the ${\frak su}(3)$ density prediction for the quadrupole deformation is expected. Note, though, that the energy of the cranked anisotropic oscillator is not proportional to $I^2$. When the oscillator frequencies are optimized, the self-consistent energy of the cranked anisotropic oscillator is
\be
E_{\rm CAO} = 3 \hbar \omega_0 \left( \Sigma_1\left( \Sigma_2 \Sigma_3 + \frac{1}{4} I^2 \right) \right)^{1/3} , 
\ee 
where $\omega_0^3 = \omega_1 \omega_2 \omega_3$ is constant.

When the deformation is large the assumption of no major-shell mixing is not valid. The anisotropic oscillator energy and the expectations of the angular momentum  and quadrupole moment can be evaluated analytically in terms of the angular velocity $\omega$  \cite{valatin}. The final results cannot be expressed as analytic functions of the angular momentum, but numerical calculations are elementary. For rotation around the 1-axis, the numerical values for the quadrupole deformations from density matrix theory, Eqs.~(\ref{k=0band},\ref{k=ilevels},\ref{middleaxisrot}), and from the exact problem of the cranked anisotropic harmonic oscillator are compared for two even-even $ds$-shell nuclei: $^{20}$Ne and $^{24}$Mg. 

The case of the highly deformed nucleus $^{20}$Ne is presented in Table \ref{Ne20cranking}. This nucleus is described by the $(8,0)$ ${\frak su}(3)$ representation, which, for collective rotation around the short 1-axis, corresponds to $\Sigma_1=\Sigma_2=14$, $\Sigma_3=22$. The deformations from the density matrix method are calculated from Eq.~(\ref{k=0band}). These are compared to the values calculated numerically with exact cranking of the anisotropic oscillator \cite{gcranking92}. The  agreement is excellent. In particular, the band end-points -- prolate spheroid at $I=0$ and noncollective oblate spheroid at $I=8$ -- are in perfect agreement.

The case of the tri axially deformed nucleus $^{24}$Mg is more interesting. The dominant ${\frak su}(3)$ representation for this nucleus is $(8,4)$, which corresponds to $\Sigma_3=28$, $\Sigma_2=20$, $\Sigma_1=16$. In Table \ref{Mg24K0cranking} the deformations from the cranked anisotropic oscillator are compared to the values from Eq.~(\ref{k=0band}) for collective rotation around its short $1$-axis. The differences between the two theories are negligible.

In Table \ref{Mg24KIcranking} the deformations from the cranked anisotropic oscillator are compared to the values from Eq.~(\ref{k=ilevels}). The agreement is excellent and shows that Eq.~(\ref{k=ilevels}) can be interpreted as the formula for the band heads: The nucleus rotates around its long axis. At $I=4$ the nucleus is a prolate spheroid rotating non-collectively. This is the maximum angular momentum at which a band occurs for the dominant representation (8,4) according to the Elliott model.

\section{Conclusion}
\label{sec:level4}

The density matrix method provides a simple geometrical interpretation for the rotational states in ${\frak su}(3)$ irreducible representations. Each density in the dual space has a direct physical interpretation as the expectation of observables in the algebra. But the quantum superposition principle and state fluctuations are not incorporated directly into the density theory because the admissible densities do not form a vector space. Nevertheless, the density formulation reproduces many properties of the quantized irreducible representations. This situation is similar to Hartree-Fock and its relationship to the quantized shell model. Although Hartree-Fock was founded on the independent fermion assumption, density matrix theory shows that, when viewed from an appropriate perspective, the essential character of the mean field method does not demand this assumption. In fact density matrix theory can be applied effectively to describe collective rotational states.

$SU(3)$ density theory is more tractable and than either the irreducible representation theory or the cranked anisotropic oscillator. The densities are solutions to a system of algebraic equations that are given immediately by the model's ansatz. In contrast, even for principal axis rotations, the cranked anisotropic oscillator  requires an extended argument involving energy minimization to attain the same conclusions. The irreducible representations, determined from the theory of highest weights, are difficult to work with in the noncanonical angular momentum basis. For example, the eigenvalues of $X_3$ in an irreducible representation are difficult to compute, yet the values from density matrix theory are a back-of-the-envelope calculation. This illustrates the power of the density matrix theory and indicates its potential for analyzing more complex algebraic models.

Density matrix theory transforms the angular momentum decomposition problem into a geometrical analysis of the range of the angular momentum function on admissible surfaces. In many cases, such as $SU(3)$, this is technically easier than the mathematical procedure of identifying irreducible subspaces of $SO(3)$ within an irreducible representation of the model's algebra ${\frak g}$. The latter is a difficult task when $SO(3)$ is not canonically embedded in ${\frak g}$.

In density matrix theory, the cubic ${\frak su}(3)$ Casimir plays a role equal to the quadratic Casimir in determining the set of admissible densities. In the usual ${\frak su}(3)$ model, the quadratic Casimir has a distinguished part because the quadrupole-quadrupole interaction is a linear combination of ${\cal C}_2$ and $I^2$.

The density theory allows for many solutions, including tilted rotation in a principal plane and more intricate rotational configurations. The physical interpretation of these solutions is simple and may be adopted for the corresponding state vectors in irreducible representations.

The density matrix method may be applied to any dynamical symmetry algebra ${\frak g}$. The admissible densities of the model are a level surface of the Casimirs in the algebra's dual space. For a semisimple Lie algebra, the dimension of the generic level surface ${\cal O}_\rho$ equals the dimension of the algebra minus the rank of the algebra. There are also singular level surfaces whose dimension is even smaller, e.g., $\lambda=0$ or $\mu=0$ level surfaces are just four dimensional. In nuclear applications the angular momentum algebra ${\frak so}(3)$ is a subalgebra of ${\frak g}$.  Hence, a rotationally invariant energy functional enables a further reduction in the dimension to $\dim {\cal O}_\rho - 4$ after rotation of the system to an intrinsic frame (a reduction by three, the dimension of the rotation group, and restriction to fixed angular momentum $I$). For the  ${\frak su}(3)$ problem, the dimension of the algebra is eight, its rank is two, and the effective dimension is two for a rotational scalar energy functional.

In future work the method will be applied to other algebras relevant to nuclear structure science, e.g. the general collective motion algebra ${\frak gcm}(3)$ corresponding to the extended Bohr-Mottelson model, which includes quantum vorticity \cite{Rose88}, the symplectic algebra ${\frak sp}(3,{\Bbb R})$ \cite{ROS77a,rowe96}, and the interacting boson model ${\frak u}(6)$ and its subalgebras ${\frak u}(5)$, ${\frak so}(6)$, and ${\frak so}(5)$ \cite{iachello}. It should be emphasized that the only restriction is that the physically relevant degrees of freedom span a Lie algebra of observables.

The Hohenberg-Kohn theorem of density functional theory was generalized to establish the existence of an energy functional for arbitrary dynamical algebras whose minimum is the density of the exact ground state \cite{GTs98}. But, like the original Hohenberg-Kohn result \cite{HK64}, this is an existence theorem for which an explicit construction of the density functional from the Hamiltonian is not known. Nevertheless, like the Hohenberg-Kohn theorem, it suggests a promising avenue of research to solve complex many-body problems. Two energy functionals for ${\frak su}(3)$ were considered in this paper, one motivated by the classical theory of rigid rotations, the other -- from the mathematics of integrity basis theory.

The surface of admissible states is, in fact, an orbit of the coadjoint action of the Lie group in the dual space \cite{vogan,Marsden1}. These orbit surfaces are equipped with a Poisson bracket and symplectic structure. In particular, each surface is always even-dimensional and admits canonical coordinates. An energy functional defines a Hamiltonian function, and, from the Poisson bracket, a Hamiltonian dynamical system. Hence the dynamics of density matrices is well-defined. Even though this paper studies equilibrium densities in the ${\frak su}(3)$ theory, normal mode and other dynamical properties may be investigated too. These are the analogues of the random phase approximation and time-dependent Hartree-Fock from conventional mean field theory.

Finally the surfaces that satisfy the generalized Bohr-Sommerfeld quantization condition may be used to construct explicit irreducible representations. The procedure to obtain the irreps is called geometric quantization \cite{Souriaufr,Souriau,kirillov1}. The method was applied in prior work to determine the irreducible representations of the rotational and Bohr-Mottelson theories \cite{georgeed79}.

\acknowledgements
We would like to thank Alan Goodman for valuable discussions and suggestions.

\appendix
\section{Dual space functions}\label{lambdamap}

One important concept in density matrix theory is that of a coordinate function. For each Lie algebra element $Z\in {\frak su}(3)$ there is a real-valued ``coordinate" function $\varphi(Z)$ defined on the dual space: the value of the function $\varphi(Z)$ at the point $\rho\in {\frak su}(3)^\ast$ is defined by 
\be
\varphi(Z) (\rho) = \langle \rho \, , Z\rangle = \mbox{tr}\, (\rho\,Z) .
\ee 
These functions separate points, i.e. if $\rho_1$ and $\rho_2$ are distinct densities, then there exists a Lie algebra element $Z$ such that $\varphi(Z) (\rho_1)\neq\varphi(Z) (\rho_2)$. The physical interpretation of $\varphi(Z)$ is that it is the real-valued function corresponding to the hermitian operator $\sigma(Z)$. The value of the observable $Z$ at the density $\rho$ is the number  $\varphi(Z) (\rho)$. 

Polynomials in the algebra generators are relevant for many physical applications, e.g., the Casimir invariants and the cubic and quartic scalars $X_3$ and $X_4$. In this appendix the extension of $\varphi$ from the domain of Lie algebra elements to the domain of polynomials is defined.

First, the domain of $\varphi$ is extended to the symmetric algebra ${\cal S}({\frak g})$ of the Lie algebra ${\frak g}$. ${\cal S}({\frak g})$ is a commutative associative algebra with elements that are symmetric polynomials of elements of ${\frak g}$ (the order of multiplying the elements of ${\frak g}$ is not important) \cite{dixmier}. The extension 
\be
\varphi\;:\; {\cal S}({\frak g})\lra C^\infty({\frak g}^\ast,\Bbb{R})
\ee
is defined in the following way: if $\{Z_1, Z_2,\ldots,Z_n \}$ is a basis for the Lie algebra ${\frak g}$ and $(\epsilon_1,\epsilon_2,\ldots\epsilon_n)$ is a set of integers, then
\be\label{lambdaonS}
\varphi(Z_1^{\epsilon_1}Z_2^{\epsilon_2}\ldots Z_n^{\epsilon_n})=\varphi(Z_1)^{\epsilon_1}\varphi(Z_2)^{\epsilon_2}\ldots \varphi(Z_n)^{\epsilon_n}.
\ee

The symmetric algebra ${\cal S}({\frak g})$ and the universal enveloping algebra ${\cal U}({\frak g})$ are related through a map $\Lambda$, called symmetrization: 
\be
\Lambda\;:\;{\cal S}({\frak g})\lra{\cal U}({\frak g}). 
\ee
Given that $X$ and $Y$ are elements of the Lie algebra ${\frak g}$, $XY$ will denote their product in ${\cal S}({\frak g})$ ($XY=YX$), and $X\cdot Y$ will denote their product in ${\cal U}({\frak g})$ ($X\cdot Y=Y\cdot X +[X,Y]$). If $\{Z_1, Z_2,\ldots,Z_n \}$ is a basis for ${\frak g}$, the symmetrization mapping is, by definition,
\be
\Lambda(Z_1^{\epsilon_1}Z_2^{\epsilon_2}\ldots Z_n^{\epsilon_n})=\frac{1}{m!}\,\sum Z_{i_{p(1)}}\cdot Z_{i_{p(2)}}\cdot\ldots\cdot Z_{i_{p(m)}}\,,
\ee
where $m=\sum_i\epsilon_i$, $(i_1,i_2,\ldots,i_m)$ is a set of $m$ integers such that exactly $\epsilon_j$ of them are equal to $j$ $(1\leq j\leq n)$, and the sum is over all permutations $p$ of the $m$ integers $(1,2,\ldots,m)$. For example, $\Lambda(Z_1Z_2)=\frac{1}{2}\,(Z_1\cdot Z_2+Z_2\cdot Z_1)$. The symmetrization $\Lambda$ is a vector space isomorphism (a canonical linear bijection) \cite{dixmier}. 

Since symmetrization is an isomorphism, its inverse is defined. Hence the extension of the domain of $\varphi$ from the Lie algebra to the enveloping algebra is given by
\be
\varphi\circ\Lambda^{-1}: {\cal U}({\frak g})\lra C^\infty({\frak g}^\ast,\Bbb{R}).
\ee

\begin{figure}
\caption{The deformation $\beta$ is a unique function of the total angular momentum $I$. At the maximum angular momentum $I=\lambda + \mu$, the deformation is a minimum $\beta_{\min} = | \lambda - \mu | /3$. \label{fig1}}
\end{figure}

\begin{figure}
\caption{The triaxiality parameter $\gamma$ as a function of the total angular momentum $I$ for rotation around the long and short principal axes when $(\lambda, \mu) = (8, 4)$. \label{fig2}}
\end{figure}

\begin{figure}
\caption{For $(\lambda,\mu)=(8,4)$ the projection $K$ onto the symmetry axis of the total angular momentum $I$ of a rotating prolate spheroid is plotted versus the angular momentum. \label{fig3}}
\end{figure}

\begin{figure}
\caption{For $(\lambda,\mu)=(8,4)$ the projection of the angular momentum on the short axis is plotted versus the total angular momentum for a sequence of triaxial tilted rotors which are critical points of energy functional I.} \label{fig4}
\end{figure}

\begin{figure}
\caption{ For $(\lambda,\mu)=(8,4)$, sequences of triaxial principal plane rotors are shown as plots of $\gamma$ and $(I_1/I)^2$ versus the angular momentum $I$ when $\nu=20/3$ for energy functional II.} \label{fig5}
\end{figure}

\begin{figure}
\caption{ For $(\lambda,\mu)=(8,4)$, sequences of triaxial principal plane rotors are shown as plots of $\gamma$ and $(I_1/I)^2$ versus the angular momentum $I$ when $\nu=0$ for energy functional II.} \label{fig6}
\end{figure}

\begin{figure}
\caption{ For $(\lambda,\mu)=(8,4)$, sequences of triaxial principal plane rotors are shown as plots of $\gamma$ and $(I_1/I)^2$ versus the angular momentum $I$ when $\nu=-4/3$ for energy functional II.} \label{fig7}
\end{figure}


\begin{table} 
\caption[]{\label{mu0} Deformations for rotation around the 1-axis, $\mu=0$}
\begin{tabular}{ccllr}
${I/\lambda}$ & ${q_1/\lambda}$ & ${q_2/\lambda}$ & ${q_3/\lambda}$ & $\gamma^\circ$ \\ \tableline
0 & $-{1\over3}$ & $-{1\over 3}$ & $+{2\over3}$  & $0.00$ \\
${1\over 4}$ & $-{1\over 3}$ & $+{1\over 6}-{\sqrt{15}\over 8}$ & $+{1\over 6}+{\sqrt{15}\over 8}$ & $0.81$ \\ 
${1\over 2}$ & $-{1\over 3}$ & $+{1\over 6}-{\sqrt{3}\over 4}$ & $+{1\over 6}+{\sqrt{3}\over 4}$ &  $3.69$ \\ 
${3\over 4}$ & $-{1\over 3}$ & $+{1\over 6}-{\sqrt{7}\over 8}$ & $+{1\over 6}+{\sqrt{7}\over 8}$ & $11.12$ \\ 
1 & $-{1\over 3}$ & $+{1\over 6}$  & $+{1\over 6}$  & $60.00$ \\ 
\end{tabular}
\end{table}

\begin{table} 
\caption[]{\label{lambda0} Deformations for rotation around the 1-axis, $\lambda=0$}
\begin{tabular}{llllr} 
$I/\mu$ & $q_1/\mu$ & $q_2/\mu$ & $q_3/\mu$ & $\gamma^\circ$\\ \tableline
0 & $+{1\over 3}$ & $+{1\over 3}$  & $-{2\over3}$  & $60.00$ \\
${1\over 4}$ & $+{1\over 3}$ & $-{1\over 6}+{\sqrt{15}\over 8}$ & $-{1\over 6}-{\sqrt{15}\over 8}$ & $59.19$ \\ 
${1\over 2}$ & $+{1\over 3}$ & $-{1\over 6}+{\sqrt{3}\over 4}$ & $-{1\over 6}-{\sqrt{3}\over 4}$ &  $56.31$ \\ 
${3\over 4}$ & $+{1\over 3}$ & $-{1\over 6}+{\sqrt{7}\over 8}$ & $-{1\over 6}-{\sqrt{7}\over 8}$ & $48.88$ \\ 
1 & $+{1\over 3}$ & $-{1\over 6}$  & $-{1\over 6}$  & $0.00$ \\
\end{tabular}
\end{table}

\begin{table}
\caption[]{ \label{Mgspheroids}Tilted rotation of spheroids for the (8,4) ${\frak su}(3)$ representation. The projection of the angular momentum on the symmetry axis is $K$.}
\begin{tabular}{cr@{.}lr@{.}lr@{.}lr@{.}l}
 & \multicolumn{4}{c}{Prolate} &\multicolumn{4}{c}{Oblate}\\
$I$ & \multicolumn{2}{c}{$q$} &  \multicolumn{2}{c}{$K$} & \multicolumn{2}{c}{$q$} & \multicolumn{2}{c}{$K$}\\
\tableline
4 & -3&33 &   4&00 & --&-- &  --&-- \\ 
5 & -3&22 &   4&04 & --&-- &  --&-- \\ 
6  & -3&07 &   4&07 & --&-- &  --&--\\ 
7 & -2&89 &   4&10 & --&-- &  --&-- \\ 
8 & -2&67 &   4&07 & 2&67 &  8&0 \\ 
9 & -2&39 &   3&96 & 2&39 &  8&3 \\ 
10 & -2&03 &   3&62 & 2&03 &  8&7 \\
11 & -1&54 &   2&34 & 1&54 &  9&4 \\ 
12 & --&-- &   --&-- & 0&67 &  12&0 \\ 
\end{tabular}
\end{table}

\begin{table}
\caption[]{\label{tiltedtriaxialband} Principal plane rotations, $I_2=0$, for the (8,4) ${\frak su}(3)$ representation}
\begin{tabular}{r@{.}lr@{.}lr@{.}lr@{.}lr@{.}lr@{.}lr@{.}lr@{.}lr@{.}l}
\multicolumn{2}{c}{$I$} & \multicolumn{2}{c}{$q^{}_1$} &  \multicolumn{2}{c}{$q^{}_2$} & \multicolumn{2}{c}{$q^{}_3$} & \multicolumn{2}{c}{$I_3$} & \multicolumn{2}{c}{$q^{}_1$} &  \multicolumn{2}{c}{$q^{}_2$} & \multicolumn{2}{c}{$q^{}_3$} & \multicolumn{2}{c}{$I_3$}\\
\tableline
4&00 & -3&33 &  -3&33 & 6&67 & 4&00 & --&-- &  --&-- & --&-- & --&--\\ 
5&00 & -3&33 &  -3&10 & 6&44 & 4&05 & --&-- &  --&-- & --&-- & --&--\\ 
6&00 & -3&33 &  -2&80 & 6&14 & 4&11 & --&-- &  --&-- & --&-- & --&--\\ 
7&00 & -3&34 &  -2&42 & 5&76 & 4&19 & --&-- &  --&-- & --&-- & --&--\\ 
8&00 & -3&35 &  -1&92 & 5&27 & 4&31 & -5&33 &   2&67 &  2&67 &  0&00\\ 
9&00 & -3&37 &  -1&24 & 4&61 & 4&48 & -4&76 &   2&07 &  2&68 &  3&47\\ 
10&00 & -3&54 &  0&06 & 3&48 & 4&80 & -3&88 &   0&93 &  2&95 &  4&78\\
10&08 & -3&68 &  0&50 & 3&18 & 4&84 & -3&68 &  0&50 & 3&18 & 4&84\\  
\end{tabular}
\end{table}

\begin{table} 
\caption[$X_3$ values for the $K=0$ band levels of (8,4)]{\label{Mg24K0X3} $X_3$ values for the ``$K=0$" band of (8,4). The values for $q_1,q_2,q_3$  are calculated from Eq.~(\ref{k=0band})}

\begin{tabular}{cr@{.}lr@{.}lr@{.}lr@{.}lr@{.}lr@{.}l}

$I$& \multicolumn{2}{c}{$q_1$} &  \multicolumn{2}{c}{$q_2$} & \multicolumn{2}{c}{$q_3$} &  \multicolumn{2}{c}{$\gamma^\circ$} & \multicolumn{2}{c}{$\varphi\left(X_3\right)$} &
\multicolumn{2}{c}{$\hat{X}_3$}\\ 

 & \multicolumn{2}{c}{} & \multicolumn{2}{c}{}& \multicolumn{2}{c}{} & \multicolumn{2}{c}{} &\multicolumn{2}{c}{density}& \multicolumn{2}{c}{eigenvalue} \\ \tableline 
 
0&  $-5$&33 &  6&67 &  $-1$&33 &  19&11 &  0&00 & 0&00\\ 
2&  $-5$&33 &  6&54 &  $-1$&21 &  20&02 &  $-21$&33 & $-24$&52\\ 
4&  $-5$&33 &  6&13 &  $-0$&80 &  23&13 &  $-85$&33 & $-93$&90\\ 
6&  $-5$&33 &  5&31 &  0&02 &  30&20 &  $-191$&98 & $-214$&23\\ 
8&  $-5$&33 &  2&67 &  2&67 &  60&00 &  $-341$&31 &  $-383$&81\\ 
\end{tabular}
\end{table}

\begin{table} 
\caption[$X_3$ values for the ``$K=4$" band levels of (8,4)]{\label{Mg24K4X3} $X_3$ values for the ``$K=4$" band of prolate spheroids for (8,4)}

\begin{tabular}{cr@{.}lr@{.}lr@{.}l}

$I$ & \multicolumn{2}{c}{$q$} &  \multicolumn{2}{c}{$\varphi\left(X_3\right)$} &
\multicolumn{2}{c}{$\hat{X}^{\Sc(3)}$}\\ 

 & \multicolumn{2}{c}{} & \multicolumn{2}{c}{density}& \multicolumn{2}{c}{eigenvalue} \\ \tableline 
 
4 &  $-3$&33 & 106&56 & 108&18\\ 
5 &  $-3$&22 & 77&17 & 72&06\\ 
6 &  $-3$&07 & 42&04 & 31&94\\ 
7 &  $-2$&89 & 4&13 & $-16$&23\\ 
8 &  $-2$&67 & $-38$&20 & $-54$&89\\ 
9 &  $-2$&39 & $-81$&20 & $-123$&64\\ 
10&  $-2$&03 & $-123$&20 & $-141$&88\\ 
11&  $-1$&54 &  $-161$&04 & $-250$&00\\ 
\end{tabular}
\end{table}

\begin{table} 
\caption[$X_3$ values for the ``$K=2$" band levels of (8,4)]{\label{Mg24K2X3} $X_3$ values for the ``$K=2$" band of (8,4). The fundamental system is solved for rotations in the 1-3 plane}

\begin{tabular}{cr@{.}lr@{.}lr@{.}lr@{.}lr@{.}lr@{.}l}

$I$ & \multicolumn{2}{c}{$q_1$} &  \multicolumn{2}{c}{$q_2$} & \multicolumn{2}{c}{$q_3$} & \multicolumn{2}{c}{$\gamma^\circ$} & \multicolumn{2}{c}{$\varphi\left(X_3\right)$} &
\multicolumn{2}{c}{$\hat{X}^{\Sc(3)}$}\\ 

 & \multicolumn{2}{c}{} & \multicolumn{2}{c}{} & \multicolumn{2}{c}{} & \multicolumn{2}{c}{} &\multicolumn{2}{c}{density}& \multicolumn{2}{c}{eigenvalue} \\ \tableline 
 
2 & $-5$&06 & $-1$&60 & 6&67 & 16&70 & 26&67 & 24&52\\ 
3 & $-5$&07 & $-1$&44 & 6&51 & 17&83 &  0&69 &  0&00\\
4 & $-5$&08 & $-1$&21 & 6&28 & 19&57 & $-35$&78 & $-14$&28\\
5 & $-5$&08 & $-0$&88 & 5&96 & 22&14 & $-82$&90 & $-72$&00\\ 
6 & $-5$&09 & $-0$&43 & 5&53 & 25&96 & $-140$&86 & $-70$&71\\
7 & $-5$&11 &  0&21 &  4&90 &  32&04 & $-210$&35 & $-183$&10\\
8 & $-5$&13 &  1&32 &  3&82 &  44&27 & $-292$&52 & $-159$&29\\ 
9 & \multicolumn{2}{c}{$\in {\Bbb C}$} &  \multicolumn{2}{c}{$\in {\Bbb C}$} & \multicolumn{2}{c}{$\in {\Bbb C}$} &  \multicolumn{2}{c}{--} & \multicolumn{2}{c}{--} & $-336$&36\\ 
10 & \multicolumn{2}{c}{$\in {\Bbb C}$} &  \multicolumn{2}{c}{$\in {\Bbb C}$} & \multicolumn{2}{c}{$\in {\Bbb C}$} &  \multicolumn{2}{c}{--} & \multicolumn{2}{c}{--} & $-291$&45\\ 
\end{tabular}
\end{table}

\begin{table} 
\caption[Deformations for $^{20}$Ne]{\label{Ne20cranking} Deformations for  $^{20}$Ne calculated with the cranked anisotropic oscillator and the density matrix method}
\begin{tabular}{cr@{.}lr@{.}lr@{.}lr@{.}lr@{.}lr@{.}l} 

$I$ & \multicolumn{4}{c}{$q_1$} &  \multicolumn{4}{c}{$q_2$} & \multicolumn{4}{c}{$q_3$} \\ 

& \multicolumn{2}{c}{cranking} & \multicolumn{2}{c}{density} & \multicolumn{2}{c}{cranking} & \multicolumn{2}{c}{density} & \multicolumn{2}{c}{cranking} & \multicolumn{2}{c}{density} \\ \tableline 

0& $-2$&67 & $-2$&67 & $-2$&67 & $-2$&67 & 5&33 & 5&33 \\ 
2& $-2$&67 & $-2$&67 & $-2$&58 & $-2$&54 & 5&25 & 5&21 \\ 
4& $-2$&69 & $-2$&67 & $-2$&31 & $-2$&13 & 5&00 & 4&80 \\  
6& $-2$&72 & $-2$&67 & $-1$&78 & $-1$&31 & 4&50 & 3&98 \\ 
8& $-2$&67 & $-2$&67 & 1&33 & 1&33 & 1&33 & 1&33 \\ 
\end{tabular}
\end{table}

\begin{table} 
\caption[Deformations for $K=0$ band of $^{24}$Mg]{\label{Mg24K0cranking} Deformations for the ``$K=0$" band of  $^{24}$Mg calculated with the cranked anisotropic oscillator and Eq.~(\ref{k=0band}) from the density matrix method}
\begin{tabular}{cr@{.}lr@{.}lr@{.}lr@{.}lr@{.}lr@{.}l} 

$I$ & \multicolumn{4}{c}{$q_1$} & \multicolumn{4}{c}{$q_2$} & \multicolumn{4}{c}{$q_3$} \\ 

& \multicolumn{2}{c}{cranking} & \multicolumn{2}{c}{density} & \multicolumn{2}{c}{cranking} & \multicolumn{2}{c}{density} & \multicolumn{2}{c}{cranking} & \multicolumn{2}{c}{density} \\ \tableline 

0& $-5$&33 & $-5$&33 & $-1$&33 & $-1$&33 &  6&67 & 6&67 \\ 
2& $-5$&34 & $-5$&33 & $-1$&23 & $-1$&21 & 6&56 & 6&54 \\ 
4& $-5$&34 & $-5$&33 & $-0$&90 & $-0$&80 & 6&25 & 6&13 \\  
6& $-5$&36 & $-5$&33 & $-0$&26 & 0&02 & 5&62 & 5&31 \\ 
8& $-5$&33 & $-5$&33 & 2&67 & 2&67 & 2&67 & 2&67 \\ 
\end{tabular}
\end{table}

\begin{table} 
\caption[Band Termination for $^{24}$Mg]{\label{Mg24KIcranking} Deformations for the band heads of $^{24}$Mg calculated with the cranked anisotropic oscillator and Eq.~(\ref{k=ilevels}) from the density matrix method}
\begin{tabular}{cr@{.}lr@{.}lr@{.}lr@{.}lr@{.}lr@{.}l} 

$I$ & \multicolumn{4}{c}{$q_1$} &  \multicolumn{4}{c}{$q_2$} & \multicolumn{4}{c}{$q_3$} \\ 

& \multicolumn{2}{c}{cranking} & \multicolumn{2}{c}{density}& \multicolumn{2}{c}{cranking}& \multicolumn{2}{c}{density} & \multicolumn{2}{c}{cranking} & \multicolumn{2}{c}{density} \\ \tableline 

0& 6&67 & 6&67 & $-5$&33 & $-5$&33 & $-1$&33 & $-1$&33 \\ 
2& 6&66 & 6&67 & $-5$&09 & $-5$&07 & $-1$&58 & $-1$&60 \\ 
4& 6&67 & 6&67 & $-3$&33 & $-3$&33 & $-3$&33 & $-3$&33 \\  
\end{tabular}
\end{table}

\bibliographystyle{prsty}
\bibliography{database}

\end{document}